\documentclass[onecolumn]{emulateapj}
\newcommand{\beq}{\begin{equation}}
\newcommand{\beqa}{\begin{eqnarray}}
\newcommand{\eeq}{\end{equation}}
\newcommand{\eeqa}{\end{eqnarray}}
\newcommand{\simg}{\gtrsim}

\shorttitle{Relativistic stars
with magnetic fields and meridional flow}
\shortauthors{Ioka \& Sasaki}
\begin{document}
\title{
Relativistic stars
with poloidal and toroidal magnetic fields and meridional flow
}
\author{
Kunihito Ioka\altaffilmark{1} and Misao Sasaki\altaffilmark{2}
}
\altaffiltext{1}{Department of Earth and Space Science,
Osaka University, Toyonaka 560-0043, Japan}
\altaffiltext{2}{Yukawa Institute for Theoretical Physics, Kyoto University,
Kyoto 606-8502, Japan}
\email{ioka@vega.ess.sci.osaka-u.ac.jp, misao@yukawa.kyoto-u.ac.jp}

\begin{abstract}
We investigate stationary axisymmetric configurations of magnetized stars
in the framework of general relativistic ideal magnetohydrodynamics.
Our relativistic stellar model
incorporates a toroidal magnetic field and meridional flow
in addition to a poloidal magnetic field 
for the first time.
The magnetic field and meridional flow are treated as perturbations,
but no other approximation is made.
We find that the stellar shape can be prolate rather than oblate
when a toroidal field exists.
We also find that, for fixed baryonic mass and total magnetic helicity,
more spherical the star is, lower the energy it has.
Further, we find two new types of the frame dragging effect
which differ from the standard one in a rotating star or Kerr geometry.
They may violate the reflection symmetry about the equatorial plane.
\end{abstract}


\section{Introduction}

Equilibrium configurations of stars are fundamental to study
their dynamics, such as the precession, oscillation
and gravitational wave emission
(Ioka 2001; Bonazzola \& Gourgoulhon 1996; Cutler 2002; 
Ioka \& Taniguchi 2000).
Among the main factors that affect the stellar structure,
the magnetic stress may be very important
especially for neutron stars
(Bocquet et al. 1995; Bonazzola \& Gourgoulhon 1996; Ioka 2001;
Konno, Obata, \& Kojima 1999).
While most neutron stars have magnetic fields of $\sim 10^{12}$-$10^{13}$\,G,
there is growing evidence for 
the existence of super-magnetized neutron stars with
$\sim 10^{14}$-$10^{15}$\,G, the so-called magnetars,
and their birthrate is estimated to be high, 
$\simg 10 \%$ of all neutron stars
(Kouveliotou et al. 1998; Duncan \& Thompson 1992;
Mereghetti et al. 2002; Thompson 2001).
Larger magnetic fields $\simg 10^{16}$\,G may be
generated by the helical dynamo inside a new born neutron star
(Duncan \& Thompson 1992; Thompson \& Duncan 1993).
Even the maximum field strength
allowed by the virial theorem $\sim 10^{18}$\,G
could be achieved 
if the central engine of gamma-ray bursts are magnetars
(Usov 1992; Nakamura 1998; Klu$\acute {\rm z}$niak \& Ruderman 1998;
Wheeler, H$\ddot {\rm o}$flich, \& Wang 2000).
In a previous paper (Ioka \& Sasaki 2003),
we presented a formalism to study 
the equilibrium configurations of magnetars.
In this paper, based on this formalism,
we calculate the actual configurations of magnetars.

There has been some work on
stationary axisymmetric configurations
of magnetized stars in the framework of 
general relativistic magnetohydrodynamics (MHD),
but allowing the existence of only a poloidal magnetic field
(Bonazzola et al. 1993; Bocquet et al. 1995; Konno, Obata, \& Kojima 1999;
Cardall, Prakash, \& Lattimer 2001).
The reason is that the existence of only a poloidal field is 
compatible with the circularity of the spacetime 
(Gourgoulhon \& Bonazzola 1993; Oron 2002),
and once the circularity is assumed, the spacetime metric
becomes substantially simple.
In a circular spacetime, there exists a family of two-surfaces
everywhere
orthogonal to the plane defined by the two Killing vectors
associated with stationarity $\eta^\mu=(\partial/\partial t)^\mu$
and axisymmetry 
$\xi^\mu=(\partial/\partial \varphi)^\mu$
(Papapetrou 1966; Carter 1969, 1973).
Thus one may choose the coordinates 
$\left(x^{\mu}\right)=\left(t,x^1,x^2,\varphi\right)$
such that the metric components $g_{01}$, $g_{02}$, $g_{31}$
and $g_{32}$ are identically zero.
As a consequence, the problem is simplified dramatically.
However non-negligible toroidal magnetic fields are likely to exist
in nature. In addition, a meridional flow may also exist in
the interior of a neutron star,
which also violates the circularity of the spacetime
(Gourgoulhon \& Bonazzola 1993; Oron 2002).
Thus, we have to consider noncircular spacetimes.

The problem to obtain an equilibrium configuration of 
a magnetized star can be separated into two parts.
The first part is the matter and electromagnetic field equations
in a given spacetime geometry (e.g., 
Zanotti \& Rezzolla 2002; Rezzolla, Ahmedov, \& Miller 2001a, b).
The second part is the Einstein equations which determine the
spacetime geometry under a given configuration of matter and
electromagnetic fields.
In Ioka \& Sasaki (2003),
we formulated the first problem, i.e.,
the equations of motion for the matter and 
electromagnetic fields in a curved spacetime.
We reduced basic equations to a single differential equation,
the so-called Grad-Shafranov (GS) equation
for the magnetic flux function
in a noncircular (i.e., the most general) stationary axisymmetric
spacetime.

In this paper, we develop a relativistic stellar model
in which both a toroidal magnetic field and meridional flow
are incorporated, in addition to a poloidal magnetic field,
in a self-consistent way for the first time.
We solve the GS equation and the perturbed
Einstein equations simultaneously to determine
the effects of toroidal fields and meridional
flows on the spacetime geometry and stellar structure.
We assume that magnetic fields are weak compared with gravity.
This assumption is valid as long as
the magnetic field strength is smaller than the maximum
value allowed by the virial theorem $\sim 10^{18}$\,G
(Bocquet et al. 1995; Bonazzola \& Gourgoulhon 1996).
We therefore treat the magnetic field as a small perturbation
on an already-known non-magnetized, non-rotating configuration.
Our approach is similar to that developed 
for slowly rotating stars
(Chandrasekhar 1933; Hartle 1967; Hartle \& Thorne 1968;
Chandrasekhar \& Miller 1974),
in which the perturbation parameter is the angular velocity,
whereas it is the magnetic flux function in our case.

This paper is organized as follows.
In \S~\ref{sec:GS}, we briefly review general relativistic 
ideal MHD in stationary axisymmetric spacetimes
and recapitulate the GS equation obtained in
Ioka \& Sasaki (2003).
In \S~\ref{sec:weak}, we take a weak magnetic field limit
of the GS equation. This makes it possible to solve
the GS equation by separation of variables.
In \S~\ref{sec:metric}, we derive the perturbed
Einstein equations for the metric perturbations.
In \S~\ref{sec:results}, we numerically solve
the GS equation and the perturbed Einstein equations 
to obtain the magnetic field and fluid flow structure,
the mass shift due to the magnetic field,
the deformation of stars, and the frame dragging effects.
Finally, we summarize our results in \S~\ref{sec:sum}.

We use the units $c=G=k_B=1$.
Greek indices ($\mu,\,\nu,\,\alpha,\,\beta,\,\cdots$) run from $0$ to 
$3$, small Latin indices ($i,\,j,\,k,\,\cdots$) from $1$ to $3$,
and capital Latin indices ($A,\,B,\,C,\,\cdots$) from $1$ to $2$,
where $x^0=t$ and $x^3=\varphi$.
The signature of the 4-metric is $(-,+,+,+)$.

\section{General relativistic ideal MHD
in stationary axisymmetric spacetimes}
\label{sec:GS}

The ideal MHD equations for a stationary axisymmetric system
can be reduced to a single equation, 
the relativistic Grad-Shafranov (GS) equation
(Ioka \& Sasaki 2003; Punsly 2001).
It is a second-order, nonlinear partial differential equation
for a quantity called the flux function $\Psi$.
The flux function $\Psi$ is such that it is constant over each surface
generated by rotating the magnetic field lines
(or equivalently the flow lines) about the axis of symmetry and 
the GS equation determines the transfield equilibrium.
Any physical quantities can be calculated from the solution $\Psi$
of the GS equation. 
The relativistic GS equation in a noncircular (i.e., the most general)
stationary axisymmetric spacetime was derived in 
Ioka \& Sasaki (2003).
In this section, we briefly review the derivation.
The weak magnetic field limit of the GS equation
is given in \S~\ref{sec:weak}.

\subsection{General relativistic MHD equations}
The basic equations for general relativistic ideal MHD 
are as follows
(Lichnerowicz 1967; Novikov \& Thorne 1973; Bekenstein \& Oron 1978).
Baryons are conserved,
\beqa
(\rho u^{\mu})_{;\mu}=0,
\label{eq:bacon}
\eeqa
where $\rho$ is the rest mass density 
(i.e., the baryon mass times the baryon number density)
and $u^{\mu}$ is the fluid 4-velocity with 
$u_{\mu} u^{\mu}=-1$.
The electromagnetic field is governed by the Maxwell equations,
\beqa
F_{[\mu \nu; \alpha]}&=&0,
\label{eq:maxwell1}
\\
{F^{\mu \nu}}_{;\nu}&=&4\pi J^{\mu},
\label{eq:maxwell2}
\eeqa
where $F_{\mu \nu}$ and $J^{\mu}$ are
the field strength tensor and the electric current 4-vector,
respectively.
Equation~(\ref{eq:maxwell1}) implies the existence of
a vector potential $A_\mu$; $F_{\mu \nu}=A_{\nu, \mu}-A_{\mu, \nu}$.
The electric and magnetic fields in the fluid rest frame are defined as
\beqa
E_{\mu}&=&F_{\mu \nu} u^{\nu},
\label{eq:E}
\\
B_{\mu}&=&-\frac{1}{2} \epsilon_{\mu \nu \alpha \beta}
u^{\nu} F^{\alpha \beta},
\label{eq:B}
\eeqa
where $\epsilon_{\mu \nu \alpha \beta}$ is the Levi-Civita 
antisymmetric unit tensor with $\epsilon_{0123}=\sqrt{-g}$.
In the ideal MHD, we assume the perfect conductivity, so that
\beqa
E_{\mu}&=&F_{\mu \nu} u^{\nu}=0.
\label{eq:ideal}
\eeqa
The equations of motion for the fluid are given by ${T^{\mu \nu}}_{;\nu}=0$,
where $T^{\mu \nu}$ is the total energy momentum tensor
of the fluid and the electromagnetic field,
\beqa
T^{\mu \nu}=(\rho+\rho\epsilon+p) u^{\mu} u^{\nu}+p g^{\mu \nu}
+\frac{1}{4\pi} \left[
\left(u^{\mu} u^{\nu}+\frac{1}{2} g^{\mu \nu}\right) B^2
-B^{\mu} B^{\nu}\right].
\label{eq:emten}
\eeqa
Here $\epsilon$ and $p$ 
are the internal energy per unit mass and pressure, respectively,
$B^2:=B^{\mu} B_{\nu}$
and we have
used the perfect conductivity condition~(\ref{eq:ideal}).
Assuming local thermodynamic equilibrium,
the first law of thermodynamics is given by
\beqa
d\epsilon=-pd\left(\frac{1}{\rho}\right) + T dS,
\label{eq:1law}
\eeqa
where $S$ and $T$ are the entropy per unit mass and the temperature.
Finally we supply the equation of state,
\beqa
p=p(\rho, S).
\label{eq:eos}
\eeqa

\subsection{Conservation laws in the axisymmetric stationary case}
We take $\eta^{\mu}=(\partial/\partial t)^\mu$
and $\xi^{\mu}=(\partial/\partial\varphi)^\mu$ so that
$x^{0}=t$ and $x^{3}=\varphi$ are the time and azimuthal coordinates
associated with the Killing vectors $\eta^{\mu}$ and $\xi^{\mu}$,
respectively. Thus all physical quantities
are independent of $t$ and $\varphi$.

In a stationary axisymmetric MHD system, there exist five conserved
quantities along each flow line constructed from
the energy-momentum tensor
(Bekenstein \& Oron 1978, 1979; Ioka \& Sasaki 2003):
 $D$, $L$, $\Omega$, $C$ and $S$.
By exploiting the gauge freedom to make
 $A_{\mu,\nu} \eta^{\nu}=A_{\mu,0}=0$ 
and $A_{\mu,\nu} \xi^{\nu}=A_{\mu,3}=0$,
we can show that the magnetic potential
$\Psi:=A_{\mu} \xi^{\mu}=A_{3}$ as well as the electric potential
$\Phi:=A_{\mu} \eta^{\mu}=A_{0}$ are constant along each flow line,
i.e., $u^{\mu} \Psi_{,\mu}=u^{\mu} \Phi_{,\mu}=0$.
Henceforth we label the flow line by $\Psi$, which we will refer
to as the flux function.
The $\Psi={\rm const.}$ surfaces are called the flux surfaces,
which are generated by rotating the magnetic field lines 
(or the flow lines) about the axis of symmetry.
Then one can show that
\beqa
F_{03}&=&0,
\label{eq:F03}
\\
F_{0A}&=&\Omega F_{A3},
\label{eq:F0A}
\\
F_{31}&=&-\Psi_{,1}=C\sqrt{-g} \rho u^{2},
\label{eq:F31}
\\
F_{23}&=&\Psi_{,2}=C\sqrt{-g} \rho u^{1},
\label{eq:F23}
\\
F_{12}&=&C\sqrt{-g} \rho (u^{3}-\Omega u^{0}),
\label{eq:F12}
\eeqa
where $\Omega(\Psi)$ and $C(\Psi)$ are conserved along each flow line
and hence are functions of the flux function $\Psi$.
It may be useful to rewrite the above equations as
\beqa
B^{\mu}=-C \rho \left[(u_0 + \Omega u_3) u^{\mu} 
+ \eta^{\mu} + \Omega \xi^{\mu} \right].
\label{eq:Bu}
\eeqa
In addition we can show that $E(\Psi)$, $L(\Psi)$ and $D(\Psi)$
are also conserved along each flow line where
\beqa
-D&=&\mu (u_0 + \Omega u_3),
\label{eq:defD}
\\
-E&=&\left(\mu+\frac{B^2}{4\pi \rho}\right) u_0 
+ C(u_0 + \Omega u_3) \frac{B_0}{4\pi},
\label{eq:defE}
\\
L&=&\left(\mu+\frac{B^2}{4\pi \rho}\right) u_3 
+ C(u_0 + \Omega u_3) \frac{B_3}{4\pi},
\label{eq:defL}
\eeqa
and 
\beqa
\mu=1+\epsilon+\frac{p}{\rho},
\label{eq:defmu}
\eeqa
is the enthalpy per unit mass.
These conserved quantities are not mutually independent but
there is a relation among them,
\beqa
D=E-\Omega L.
\eeqa
Except for the entropy per unit mass $S(\Psi)$, 
there are no perfectly relevant physical
interpretations of these quantities.
Nevertheless, by considering several limiting cases,
we may associate them with terms that describe their qualitative nature. 
We may call $D(\Psi)$ the fluid energy per unit mass,
$E(\Psi)$ the total energy per unit mass,
$L(\Psi)$ the total angular momentum per unit mass, 
$\Omega(\Psi)$ the angular velocity, 
and $C(\Psi)$ the magnetic field strength relative
to the magnitude of meridional flow.
Since these conserved quantities are essentially
the first integrals of the equations of motion, specification of
these functions characterizes the configuration of the electromagnetic field
and fluid flow.

\subsection{$(2+1)+1$ formalism}
To describe the metric of the noncircular (the most general) 
stationary axisymmetric spacetime in a covariant fashion,
we adopt the $(2+1)+1$ formalism
developed by Gourgoulhon \& Bonazzola (1993).
Note that this formalism is different from the $(2+1)+1$ formalism
by Maeda, Sasaki, Nakamura, \& Miyama (1980)
and by Sasaki (1984; see also Nakamura, Oohara, \& Kojima 1987),
which is suitable to the axisymmetric gravitational collapse.
Here we adopt the formalism by Gourgoulhon \& Bonazzola (1993) because 
it is more convenient for a spacetime which is not only
axisymmetric but also stationary.

Let $n^{\mu}$ be the unit timelike 4-vector orthogonal to 
the $t={\rm const.}$ hypersurface $\Sigma_t$
and oriented in the direction of increasing $t$,
\beqa
n_{\mu}=-N t_{,\mu}.
\label{eq:defn}
\eeqa
%The lapse function $N$ is determined by the requirement
%\beqa
%n_{\mu} n^{\mu}=-1.
%\eeqa
The 3-metric induced by $g_{\mu \nu}$ on $\Sigma_t$ is given by
\beqa
h_{\mu \nu}=g_{\mu \nu}+n_{\mu} n_{\nu}.
\eeqa

Similarly, let $m_{\mu}$ be the unit spacelike 4-vector orthogonal to the 
$t={\rm const.}$ and $\varphi={\rm const.}$ hypersurface $\Sigma_{t\varphi}$
and oriented in the direction of increasing $\varphi$,
\beqa
m_{\mu}=M {h_{\mu}}^{\nu} \varphi_{,\nu}=M \varphi_{|\mu},
\label{eq:defm}
\eeqa
where the vertical stroke $|$ denotes the covariant derivative
associated with the 3-metric $h_{\mu \nu}$.
%The coefficient $M$ is determined by 
%\beqa
%m_{\mu} m^{\mu}=1.
%\eeqa
The induced 2-metric on $\Sigma_{t\varphi}$ is given by 
\beqa
H_{\mu \nu}=h_{\mu \nu}-m_{\mu} m_{\nu}
=g_{\mu \nu}+n_{\mu} n_{\nu}-m_{\mu} m_{\nu}.
\eeqa
The covariant derivative associated with the 2-metric $H_{\mu \nu}$
is denoted by a double vertical stroke $\parallel$.
There is a relation between the determinants as
$\sqrt{-g}=N\sqrt{h}=N M \sqrt{H}$.

Any 4-vector can be decomposed into its projection onto $\Sigma_{t\varphi}$,
the component parallel to $n_{\mu}$ and that to $m_{\mu}$.
The Killing vectors are decomposed as
\beqa
\eta^{\mu}&=&N n^{\mu}-N^{\mu}
=N n^{\mu} - M N^{\varphi} m^{\mu} - {N_{\Sigma}}^{\mu},
\label{eq:eta}
\\
\xi^{\mu}&=&M m^{\mu} - {M_{\Sigma}}^{\mu},
\label{eq:xi}
\eeqa
where the shift vector $N^{\mu}$ is (minus) the projection
of $\eta^{\mu}$ onto $\Sigma_{t}$, 
$M_{\Sigma}{}^{\mu}$ is (minus) the projection
of $\xi^{\mu}$ onto $\Sigma_{t\varphi}$, 
and $N_{\Sigma}{}^{\mu}$ is the projection
of $N^{\mu}$ onto $\Sigma_{t\varphi}$.
For our choice of the 
coordinates, i.e., for $x^0=t$ and $x^3=\varphi$,
the component expressions for $n^{\mu}$ and $m^{\mu}$ are 
%the component expressions for $n^{\mu}$, $n_{\mu}$,
% $m^{\mu}$ and $m_{\mu}$ are 
\beqa
%n_{\mu}&=&(-N,0,0,0),
%\\
n^{\mu}&=&\left(\frac{1}{N},\frac{N^{1}}{N},
\frac{N^{2}}{N},\frac{N^{\varphi}}{N}\right),
\label{eq:ncom}
\\
%m_{\mu}&=&(-M N^{\varphi},0,0,M),
%\\
m^{\mu}&=&\left(0,\frac{{M_{\Sigma}}^{1}}{M},
\frac{{M_{\Sigma}}^{2}}{M},\frac{1}{M}\right).
\label{eq:mcom}
\eeqa
Note that $N_{\Sigma}{}^{\mu}=
\left(0,N_{\Sigma}{}^{1},N_{\Sigma}{}^{2},0\right)$
and $N^{A}=N_{\Sigma}{}^{A}+N^{\varphi} M_{\Sigma}{}^{A}$.
We can express the 4-metric $g_{\mu\nu}$ 
in terms of $N$, $N^{\varphi}$, $N_{\Sigma}{}^{A}$,
$M$, $M_{\Sigma}{}^{A}$ and $H_{AB}$ as
\beqa
g_{\mu \nu} dx^{\mu} dx^{\nu}
&=&-\left[N^2-M^2 \left(N^{\varphi}\right)^2-
N_{\Sigma}{}_{A} N_{\Sigma}{}^{A}\right] dt^2
-2 \left(M^2 N^{\varphi}-N_{\Sigma}{}^{A} M_{\Sigma}{}_{A}\right) dt d\varphi
\nonumber\\
&-&2 N_{\Sigma}{}_{A} dt dx^A
+H_{AB} dx^{A} dx^{B}
-2 M_{\Sigma}{}_{A} d\varphi dx^A
+\left(M^2+M_{\Sigma}{}_{A} M_{\Sigma}{}^{A}\right) d\varphi^2,
\label{eq:metric}
\eeqa
where the functions $N$, $N^{\varphi}$, $N_{\Sigma}{}^{A}$,
$M$, $M_{\Sigma}{}^{A}$ and $H_{AB}$ depend only
on the coordinate $(x^{1}, x^{2})$.
We note that the presence of $N_{\Sigma}{}^\mu$ and $M_{\Sigma}{}^\mu$
characterizes the degree of noncircularity of the spacetime.
Since we only assume that physical quantities
are independent of $x^{0}=t$ and $x^{3}=\varphi$,
the metric $g_{\mu \nu}$ in equation (\ref{eq:metric})
has some degrees of freedom in the choice of coordinates.
Nevertheless, the norms $N_{\Sigma}{}^\mu N_{\Sigma}{}_\mu$ and
$M_{\Sigma}{}^\mu M_{\Sigma}{}_\mu$ cannot be set equal to zero
for a general noncircular spacetime.
In \S~\ref{sec:GScov}, the GS equation will be given
as an equation projected onto $\Sigma_{t\varphi}$.

\subsection{Physical quantities from flux function $\Psi$}
Provided that the metric $g_{\mu\nu}$ is given and
the conserved quantities $E(\Psi)$ (or $D(\Psi)$), $L(\Psi)$, $\Omega(\Psi)$, 
$C(\Psi)$ and $S(\Psi)$ are given as functions of $\Psi$,
all the physical quantities can be evaluated once the
(effectively 2-dimensional) configuration of the flux function
$\Psi$ is known (Ioka \& Sasaki 2003).

The fluid 4-velocity is expressed as
\beqa
u^{\mu}=u_{\eta} (\eta^{\mu} + \Omega \xi^{\mu})
+u_{\xi} (\xi^{\mu} + \Theta \eta^{\mu})+\tilde u_{\Sigma}{}^{\mu},
\label{eq:uinv}
\eeqa
where 
\beqa
u_{\eta}&=&\frac{E-\Omega L}{G_{\eta} \mu}-\frac{\tilde N_{\Sigma}}{G_{\eta}}
=\frac{D}{G_{\eta} \mu}-\frac{\tilde N_{\Sigma}}{G_{\eta}},
\label{eq:uomega}
\\
u_{\xi}&=&-\frac{(L-\Theta E)}
{G_{\xi} \mu}
\left(\frac{4\pi \mu}{G_{\eta} C^2 \rho}\right)
\left(1-\frac{4\pi \mu}{G_{\eta} C^2 \rho}\right)^{-1}
+\frac{\tilde M_{\Sigma}}{G_{\xi}},
\label{eq:utheta}
\\
\tilde u_{\Sigma}{}^{\mu}&=&\frac{1}{N M C \rho} 
\epsilon^{\mu \nu} \Psi_{,\nu},
\label{eq:tuperp}
\eeqa
and we define
\beqa
\Theta&=&-\frac{\xi_{\nu}(\eta^{\nu}+\Omega \xi^{\nu})}
{\eta_{\mu}(\eta^{\mu}+\Omega \xi^{\mu})}
=-\frac{g_{03}+\Omega g_{33}}{g_{00}+\Omega g_{03}},
\label{eq:Theta}
\\
G_{\eta}&=&-(\eta_{\mu} + \Omega \xi_{\mu})(\eta^{\mu} + \Omega \xi^{\mu})
=-(g_{00}+2\Omega g_{03}+\Omega^{2} g_{33}),
\label{eq:Geta}
\\
G_{\xi}&=&(\xi_{\mu} + \Theta \eta_{\mu})(\xi^{\mu} + \Theta \eta^{\mu})
=g_{33}+2\Theta g_{03}+\Theta^{2} g_{00},
\label{eq:Gxi}
\\
\tilde N_{\Sigma}&=&\tilde u_{\Sigma}{}^{\mu}
\left(N_{\Sigma}{}_{\mu}+\Omega M_{\Sigma}{}_{\mu}\right),
\\
\tilde M_{\Sigma}&=&\tilde u_{\Sigma}{}^{\mu}
\left(M_{\Sigma}{}_{\mu}+\Theta N_{\Sigma}{}_{\mu}\right),
\\
\epsilon^{\mu \nu}&=&\epsilon^{\mu \nu \alpha \beta} n_{\alpha} m_{\beta}.
\label{eq:epsilon}
\eeqa
In equation (\ref{eq:uinv}), the orthogonality $(\eta^{\mu}+\Omega \xi^{\mu})
(\xi_{\mu}+\Theta \eta_{\mu})=0$ is satisfied.
Note that ${M_{\rm Alf}}^2:=4\pi \mu/G_{\eta} C^2 \rho$
is the square of the effective Alfv$\acute{\rm e}$n Mach number $M_{\rm Alf}$.
At the Alfv$\acute{\rm e}$n point $M_{\rm Alf}=1$,
the numerator $L-\Theta E$ in equation (\ref{eq:utheta})
should vanish to keep the velocity
$u_{\xi}$ finite.

Thus, from equations (\ref{eq:uinv}) -- (\ref{eq:tuperp}), 
given the metric $g_{\mu\nu}$ and
the conserved functions $E(\Psi)$ (or $D(\Psi)$), $L(\Psi)$, $\Omega(\Psi)$, 
$C(\Psi)$ and $S(\Psi)$, 
if the density $\rho$ and the enthalpy $\mu$
are additionally known (see below),
the fluid 4-velocity $u^{\mu}$ can be obtained from the 
flux function $\Psi$ and its first derivatives $\Psi_{,A}$.
The magnetic field is also calculated from the flux function $\Psi$
by using equation (\ref{eq:Bu}).
The $(2+1)+1$ decomposition of the fluid 4-velocity 
is performed with equations (\ref{eq:eta}), (\ref{eq:xi}) 
and (\ref{eq:uinv}) as
\beqa
u^{\mu}=u_{n} n^{\mu}+u_{m} m^{\mu}+ {u_{\Sigma}}^{\mu},
\eeqa
where 
\beqa
u_{n}&=&N (u_{\eta} + \Theta u_{\xi}),
\\
u_{m}&=&M \left[\left(\Omega-N^{\varphi}\right) u_{\eta}
+\left(1-N^{\varphi}\Theta\right) u_{\xi}\right],
\\
{u_{\Sigma}}^{\mu}&=&\tilde u_{\Sigma}{}^{\mu}
-(u_{\eta}+\Theta u_{\xi}) {N_{\Sigma}}^{\mu}
-(u_{\xi} + \Omega u_{\eta}) {M_{\Sigma}}^{\mu}.
\label{eq:uperp}
\eeqa

The pressure $p$, the internal energy $\epsilon$, 
the enthalpy $\mu$ and the temperature $T$
are functions of the density $\rho$ and the entropy $S$
from equations (\ref{eq:defmu}), (\ref{eq:1law}) and (\ref{eq:eos}).
Hence, given $S$ as a function of $\Psi$, the only remaining quantity
to be known is the density $\rho$.
The density $\rho$ is determined by 
the normalization of the 4-velocity, which yields
%\[
%u_{\mu} u^{\mu} = -G_{\eta} {u_{\eta}}^2
%+G_{\xi} {u_{\xi}}^2+
%\tilde u_{\Sigma}{}_{A}\tilde u_{\Sigma}{}^{A}
%+2 u_{\eta} \tilde u_{\Sigma}{}^{\mu}\left(\eta_{\mu}+\Omega \xi_{\mu}\right)
%+2 u_{\xi} \tilde u_{\Sigma}{}^{\mu}\left(\xi_{\mu}+\Theta \eta_{\mu}\right)
%=-1\,,
%\]
% that is,
\beqa
-\frac{D^2}{G_{\eta} \mu^2}
+\frac{(4\pi)^2 (L-\Theta E)^2}
{{G_{\xi} G_{\eta}}^2 C^4 \rho^2}
\left(1-\frac{4\pi \mu}{G_{\eta} C^2 \rho}\right)^{-2}
+\frac{H^{A B} \Psi_{,A} \Psi_{,B}}{N^2 M^2 C^2 \rho^2}
+\frac{\tilde N_{\Sigma}{}^{2}}{G_{\eta}}
-\frac{\tilde M_{\Sigma}{}^{2}}{G_{\xi}}=-1.
\label{eq:uu1}
\eeqa
This equation is called the wind equation.

The following components of the electric current are also 
calculated from the flux function,
\beqa
J^{0}&=&\frac{1}{4\pi N M \sqrt{H}}\left(N M \sqrt{H} F^{0A}\right)_{,A}
=\frac{1}{4\pi N M}\left(N M F^{0A}\right)_{||A},
\label{eq:J0}
\\
J^{3}&=&\frac{1}{4\pi N M \sqrt{H}}\left(N M \sqrt{H} F^{3A}\right)_{,A}
=\frac{1}{4\pi N M}\left(N M F^{3A}\right)_{||A},
\label{eq:J3}
\eeqa
where
\beqa
F^{0A}&=&\left(g^{00} g^{AB} - g^{0B} g^{A0}\right) \Omega \Psi_{,B}
+\left(g^{0B} g^{A3} - g^{03} g^{AB}\right) \Psi_{,B}
+\left(g^{01} g^{A2} - g^{02} g^{A1}\right) F_{12}
\nonumber\\
&=&-\frac{1}{N^2} \left[H^{AB}+
\frac{M_{\Sigma}^A M_{\Sigma}^B}{M^2}\right] \Omega \Psi_{,B}
+\frac{1}{N^2} \left[N^{\varphi} H^{AB} - 
\frac{M_{\Sigma}^A {N_{\Sigma}}^{B}}{M^2}\right] \Psi_{,B}
\nonumber\\
&-&\left[\left(\frac{{N_{\Sigma}}^{B} + N^{\varphi} M_{\Sigma}^{B}}{N^2}\right)
{\epsilon_B}^A
+\frac{M_{\Sigma}^A {N_{\Sigma}}^B M_{\Sigma}^C \epsilon_{BC}}{N^2 M^2}
\right]
\frac{F_{12}}{\sqrt{H}},
\label{eq:Fu0A}
\\
F^{3A}&=&\left(g^{30} g^{AB} - g^{3B} g^{A0}\right) \Omega \Psi_{,B}
+\left(g^{3B} g^{A3} - g^{33} g^{AB}\right) \Psi_{,B}
+\left(g^{31} g^{A2} - g^{32} g^{A1}\right) F_{12}
\nonumber\\
&=&-\frac{1}{N^2} \left[N^{\varphi} H^{AB}
-\frac{{N_{\Sigma}}^{A} M_{\Sigma}^{B}}{M^2}\right] \Omega \Psi_{,B}
-\left[\left(\frac{1}{M^2}-\left(\frac{N^{\varphi}}{N}\right)^2\right)
H^{AB} - \frac{{N_{\Sigma}}^{A} {N_{\Sigma}}^{B}}{N^2 M^2}\right]
\Psi_{,B}
\nonumber\\
&+&\left[
\left\{
\left(\frac{1}{M^2}-\left(\frac{N^{\varphi}}{N}\right)^2\right) M_{\Sigma}^{B} 
-\frac{N^{\varphi} {N_{\Sigma}}^{B}}{N^2} 
\right\}{\epsilon_{B}}^{A}
+\frac{{N_{\Sigma}}^{A} {N_{\Sigma}}^{B} M_{\Sigma}^{C} \epsilon_{BC}}{N^2 M^2}
\right]
\frac{F_{12}}{\sqrt{H}},
\label{eq:Fu3A}
\\
F_{12}&=&C \sqrt{-g} \rho (u^3 - \Omega u^0)
=C \sqrt{-g} \rho u_{\xi}
\left(1 - \Omega \Theta \right).
\eeqa
Thus, $J^0$ and $J^3$ are expressed in terms of $\Psi$ and its
first and second derivatives.

\subsection{Relativistic Grad-Shafranov equation}\label{sec:GScov}
The equation for the flux function $\Psi$, that is,
the GS equation is given by (Ioka \& Sasaki 2003)
\beqa
%&&
J^{3}-\Omega J^{0}
+\frac{1}{N M C} \epsilon^{AB} (\mu {u_{\Sigma}}_{A})_{||B}
%\nonumber\\
%&&\quad
-\rho (u_{\eta} + \Theta u_{\xi})
\left[{\frac{dE}{d\Psi}}-\Lambda {\frac{d(C \Omega)}{d\Psi}}\right]
+\rho (u_{\xi} + \Omega u_{\eta})
\left[\frac{dL}{d\Psi}-\Lambda \frac{dC}{d\Psi}\right]
+\rho T \frac{dS}{d\Psi}=0,
\label{eq:GS}
\eeqa
where %the prime denotes differentiation with respect to $\Psi$, 
the double vertical stroke $\parallel$ denotes
covariant differentiation with respect to the 2-metric $H_{AB}$, 
and we have defined the auxiliary quantity
\beqa
\Lambda=\frac{1}{4\pi}
(u_0 B_3 - u_3 B_0)=
\frac{1}{4\pi} C \rho \left(G_{\xi} u_{\xi}-\tilde M_{\Sigma}\right)
\left(g_{00}+\Omega g_{03}\right).
\label{eq:Lambda}
\eeqa
In the previous subsection, we have seen that
$u_{\eta}$, $u_{\xi}$, $u_{\Sigma}{}^{A}$, $\Theta$, 
$\rho$, $\mu$, $T$, $J^0$ and $J^3$
are all expressed in terms of $\Psi$ and its derivatives, 
given the conserved functions
$E(\Psi)$ (or $D(\Psi)$), $L(\Psi)$, $\Omega(\Psi)$, $C(\Psi)$ and $S(\Psi)$,
and the metric $g_{\mu\nu}$. 
Thus, in order to obtain the configuration of matter and 
electromagnetic fields in a curved spacetime,
we have only to solve
the GS equation~(\ref{eq:GS}) 
with the aid of the wind equation~(\ref{eq:uu1}).

\section{Grad-Shafranov equation in the weak magnetic field limit}
\label{sec:weak}

It is formidable to solve the GS equation~(\ref{eq:GS}) 
in a general case because of its high non-linearity.
However, it becomes tractable if we take the weak magnetic field limit.
In this section, we derive the GS equation in the weak field limit
with the flux function $\Psi$ being the perturbation parameter.
This perturbation method is essentially 
similar to that developed by Chandrasekhar (1933; Chandrasekhar \& Miller 1974)
and Hartle (1967; Hartle \& Thorne 1968)
for slowly rotating stars,
in which the perturbation parameter is the angular velocity.

\subsection{Zeroth order}
In Ioka \& Sasaki (2003),
we showed that no magnetic field limit
is obtained by letting $\Psi \to 0$ and $C\to \infty$.
In this limit, the GS equation~(\ref{eq:GS}) 
and the wind equation~(\ref{eq:uu1}) are reduced
to the integrability condition 
and the Bernouilli's equation for the rotating 
isentropic ($dS/d\Psi=0$) star, respectively (Ioka \& Sasaki 2003).
Thus a rotating fluid star is 
the zeroth order configuration in the weak magnetic field approximation.

Let us further suppose that there is no rotation $\Omega=0$ 
at the zeroth order in $\Psi$.
Then the background metric $\bar g_{\mu \nu}$
is spherically symmetric,
\beqa
\bar g_{\mu \nu} dx^{\mu} dx^{\nu}=-e^{2\nu} dt^2+e^{2\lambda} dr^2+
r^2 \left(d\theta^2+\sin^2\theta d\varphi^2\right),
\label{eq:backmetric}
\eeqa
and the zeroth order configuration is obtained by
the Tolman-Oppenheimer-Volkoff equations,
\beqa
m'&=&4\pi r^2 \rho (1+\epsilon),
\\
p'&=&-\frac{1}{r^2} e^{2\lambda} \rho \mu \left(m+4\pi p r^3\right),
\\
\nu'&=&-\frac{1}{\rho\mu} p',
\label{eq:nu'}
\eeqa
where the prime denotes differentiation with respect to $r$, and
the mass $m(r)$ inside radius $r$ is defined by
\beqa
e^{2\lambda}=\left(1-\frac{2 m}{r}\right)^{-1}.
\eeqa
The value $r=R_{*}$ at which $p=0$ is the radius of the star,
and the total mass of the star is given by $m(R_{*})=M_{*}$.
The boundary conditions are $m(r=0)=0$
and $2 \nu(r=R_{*})=\ln (1-2M_{*}/R_{*})$.
%Note that there are relations
%\beqa
%\nu''+\left(\nu'\right)^2-\nu'\lambda'+
%\left(\nu'-\lambda'\right)/r=8\pi p e^{2\lambda},
%\nonumber
%\eeqa
%which will be useful in \S~\ref{sec:int}.

\subsection{First order}\label{sec:1order}
We perform linearization with respect to the flux function $\Psi$.
To proceed,
we have to specify the functional forms of the conserved functions
$D(\Psi)$ (or $E(\Psi)$), $L(\Psi)$, $\Omega(\Psi)$, $C(\Psi)$ and $S(\Psi)$,
which characterize the configuration of the fluid flow 
and the electromagnetic field.
In this paper, we assume the following.

\begin{itemize}
\item $C(\Psi)$:
We consider the case in which the magnetic field is
confined in the interior of a star.
Hence, $\Psi$ must vanish at the stellar surface.
Then, the Alfv$\acute{\rm e}$n point
 $M_{\rm Alf}^2=4\pi\mu/G_{\eta} C^2 \rho=1$
exists near the stellar surface $\rho\sim 0$ in equation (\ref{eq:utheta})
unless $C$ diverges at the stellar surface.
Therefore, we assume the form
\begin{eqnarray}
C=\frac{\tilde C}{\Psi}\,,
\label{eq:Cform}
\end{eqnarray}
where $\tilde C={\rm const.}=O(1)$.
We note that as long as the magnitude of $\tilde C$ is much
greater than $\Psi$, the discussion below is valid, but
we assume it to be $O(1)$ for the sake of order counting.

\item $\Omega(\Psi)$:
Since our primary interest is the effect of magnetic fields on the
stellar structure, we consider a slow (rigid) rotation
which is of the same order as the one
induced by the presence of a magnetic field.
Hence we set
\begin{eqnarray}
\Omega={\rm const.}=O(\Psi^2).
\label{eq:Omform}
\end{eqnarray}
Here, although we may consider the case $\Omega =O(\Psi)$,
we choose not to do so, and assume that the rotation is subdominant.
Note that the observed magnetars have long rotation periods ($\sim$ sec),
so that this assumption is justified.

\item $S(\Psi)$:
We consider an isentropic star for simplicity,
\begin{eqnarray}
S={\rm const.}.
\label{eq:Sform}
\end{eqnarray}

\item $D(\Psi)$ and $L(\Psi)$:
These functions are chosen in such a way that the GS equation becomes separable
with respect to $r$ and $\theta$.
Namely,
\begin{eqnarray}
D&=&D_0+D_1 \Psi+D_2,
\label{eq:Dform}
\\
L&=&{\rm const.}=O\left(\tilde C\right),
\label{eq:Lform}
\end{eqnarray}
where $D_{0}={\rm const.}=O(1)$, $D_{1}={\rm const.}=O(\Psi)$
and $D_{2}={\rm const.}=O(\Psi^2)$.
Note that the conserved function $L(\Psi)$ does not necessarily
vanish at zeroth order at which there is no rotation.
This is because $L(\Psi)$ does not exactly coincide with
the angular momentum.

\end{itemize}
With these assumptions for the conserved quantities,
we can estimate the orders of the variables 
from equations (\ref{eq:uomega}) -- (\ref{eq:epsilon}), 
(\ref{eq:uperp}) and (\ref{eq:Bu}) as
\beqa
%&&
\Theta=O\left(\Psi^2\right),
\quad
u_{\eta}=O(1),
\quad
u_{\xi} = O\left(\Psi^2\right),
%\nonumber\\
%&&
\quad
\tilde u_{\Sigma}{}^{\mu} \sim u_{\Sigma}{}^{\mu} 
= O\left(\Psi^2\right),
\quad
B_{0} = O\left(\Psi^3\right),
\quad
B_{i} = O\left(\Psi\right),
\label{eq:order1}
\eeqa
where we have used the fact that the metric perturbations are 
$O\left(\Psi^2\right)$, i.e., in equation (\ref{eq:metric}) we have
\beqa
N^{\varphi} =O\left(\Psi^2\right),
\quad
N_{\Sigma}{}^{\mu} = O\left(\Psi^2\right),
\quad
M_{\Sigma}{}^{\mu} = O\left(\Psi^2\right),
\label{eq:order2}
\eeqa
and $g_{03}=O\left(\Psi^2\right)$, which will be
confirmed in \S~\ref{sec:metric}.
Note that $N=e^{\nu}$, $M=r\sin\theta$, $H_{rr}=e^{2\lambda}$
and $H_{\theta\theta}=r^2$ at zeroth order
from equations (\ref{eq:metric}) and (\ref{eq:backmetric}).

Now let us derive the GS equation in the weak magnetic field limit.
To $O(\Psi)$, equation (\ref{eq:GS}) is reduced to
\beqa
J^3-\rho u_{\eta} \frac{dE}{d\Psi}-\rho u_{\xi} \Lambda \frac{dC}{d\Psi}=0,
\label{eq:GS1st}
\eeqa
where, from equations (\ref{eq:J3}) and (\ref{eq:Fu3A}),
the electric current is given by
\beqa
J^{3}=\frac{1}{4\pi \sqrt{-g}} \left(\sqrt{-g} F^{3A}\right)_{,A}
=-\frac{1}{4\pi \sqrt{-\bar g}} 
\left(\sqrt{-\bar g} \frac{1}{M^{2}} H^{AB} \Psi_{,B}\right)_{,A}.
\label{eq:J31st}
\eeqa
Up to $O\left(\Psi^2\right)$,
equations (\ref{eq:Geta}), (\ref{eq:Gxi}), (\ref{eq:uomega})
and (\ref{eq:utheta}) are reduced, respectively, to
\beqa
G_{\eta}&=&-g_{00},
\\
G_{\xi}&=&g_{33},
\\
u_{\eta}&=&\frac{D}{(-g_{00}) \mu},
\\
u_{\xi}&=&\frac{4\pi L}{g_{00} g_{33} C^2 \rho}\,.
\eeqa
Hence the wind equation~(\ref{eq:uu1}) is reduced to
\beqa
\frac{D^2}{(-g_{00}) \mu^2}
=\frac{D}{\mu} u_{\eta}
=(-g_{00}) u_{\eta}^2=1,
\label{eq:wind2nd}
\eeqa
where we note that $g_{00}$ and $g_{33}$ include perturbations
up to $O(\Psi^2)$.
Then, at $O\left(\Psi\right)$,
the second term in equation (\ref{eq:GS1st}) is given by
\beqa
-\rho u_{\eta} \frac{dE}{d\Psi}=-\rho u_{\eta} 
\frac{d(D+\Omega L)}{d\Psi}=
-\rho u_{\eta} D_{1}=-\rho e^{-\nu}D_{1}\,,
\label{eq:2term1st}
\eeqa 
where we have used $u_\eta=1/\sqrt{-\bar g_{00}}=e^{-\nu}$ at zeroth order
that follows from equation (\ref{eq:wind2nd}).
Note that we have $D_0=\mu e^{\nu}$ at zeroth order. 
Similarly, from equations (\ref{eq:Lambda}), (\ref{eq:Cform}) 
and (\ref{eq:utheta}),
the third term in equation (\ref{eq:GS1st}) is approximated as
\beqa
-\rho u_{\xi} \Lambda \frac{dC}{d\Psi}=
-\frac{1}{4\pi r^2 \sin^2\theta} \left(\frac{4\pi L}{\tilde C}\right)^2 
e^{-2\nu} \Psi.
\label{eq:3term1st}
\eeqa
Then, by combining equations (\ref{eq:GS1st}), (\ref{eq:J31st}), 
(\ref{eq:2term1st}) and (\ref{eq:3term1st}),
the GS equation valid to $O(\Psi)$ is given by
\beqa
%&&
e^{\lambda-\nu}
\left[\partial_r\left(e^{\nu-\lambda}\partial_r \Psi\right)
+\frac{e^{\nu+\lambda}}{r^2}\sin\theta
\partial_\theta\left(\frac{1}{\sin\theta} 
\partial_\theta\Psi\right)\right]
%\nonumber\\
%&&\qquad
+\left(\frac{4\pi L}{\tilde C}\right)^2
e^{2(\lambda-\nu)} \Psi
+4\pi r^2 \sin^2 \theta \rho e^{2\lambda-\nu}D_1=0,
\label{eq:GSPsi}
\eeqa
where we have used the form of the
background metric $\bar g_{\mu \nu}$ given in equation (\ref{eq:backmetric}).
Note that, under the assumptions for the conserved quantities we have made,
the GS equation at this order is an inhomogeneous linear differential
equation for $\Psi$
with the last term proportional to $D_1$ as a source.

Let us separate the angular variables in the GS equation.
We expand the flux function $\Psi$ 
by the vector spherical harmonics
(Regge \& Wheeler 1957; Zerilli 1970)
as
\beqa
\Psi=\sum_{\ell=1}^{\infty} \psi_{\ell}(r) \sin\theta 
\frac{\partial P_{\ell}(\cos\theta)}{\partial \theta},
\eeqa
where $P_{\ell}$ is the Legendre polynomial of degree $\ell$,
and we have discarded the $\varphi$-dependence of the harmonics
because of the axisymmetry.
Substituting this form into equation (\ref{eq:GSPsi}),
we find the equation for the dipole $\psi_1$ as
\beqa
\psi_{1}''+\left(\nu'-\lambda'\right) \psi_{1}'
+\left(\tilde L^2 e^{-2\nu}-\frac{2}{r^2}\right) e^{2\lambda} \psi_{1}
-4\pi r^2 \rho e^{2\lambda-\nu} D_{1}=0\,,
\label{eq:psi1}
\eeqa
and for the higher multipoles as
\beqa
\psi_{\ell}''+\left(\nu'-\lambda'\right) \psi_{\ell}'
+\left(\tilde L^2 e^{-2\nu}-\frac{\ell(\ell+1)}{r^2}\right)
 e^{2\lambda} \psi_{\ell}=0\quad(\ell\geq2),
\label{eq:psiell}
\eeqa
where we have introduced
\beqa
\tilde L=\frac{4\pi L}{\tilde C}\,.
\eeqa
Equations (\ref{eq:psi1}) and (\ref{eq:psiell}) are the basic equations
for the matter and electromagnetic fields
for the conserved quantities of the assumed forms
in the weak magnetic field limit.

Since we are interested in magnetic fields confined in the
stellar interior,
or interior magnetic fields with magnitude
much larger than exterior magnetic fields,
we assume that the magnetic field vanishes at the stellar surface.
Thus, to the lowest order in $\Psi$, we require
\beqa
\psi_{\ell}(R_{*})=0,\quad \psi_{\ell}'(R_{*})=0\quad(\ell\geq1).
\label{eq:psibound}
\eeqa
Near the origin $r=0$, the regular solutions of equations (\ref{eq:psi1})
and (\ref{eq:psiell}) should have the behavior
\beqa
\psi_\ell \to r^{\ell+1} \alpha_{\ell}\quad(\ell\geq1),
\label{eq:psi1reg}
\eeqa
where $\alpha_{\ell}$ is a constant.

First consider the dipole $\psi_1$.
With the boundary conditions given above,
equation (\ref{eq:psi1}) becomes an eigenvalue equation with $\tilde L$ 
being the eigenvalue to be determined.
There will be a discrete set of eigenvalues.
For a given eigenvalue, the constant $D_{1}$ 
then just determines the normalization of $\psi_{1}$.
Now we consider the higher multipoles $\psi_\ell$ ($\ell\geq2$).
Once $\tilde L$ is determined, it is in general impossible to 
find non-trivial solutions for $\ell\geq2$ that also satisfy
the same boundary conditions~(\ref{eq:psibound}) and (\ref{eq:psi1reg}).
Thus, under our assumptions that the conserved 
quantities are of the forms~(\ref{eq:Cform}) -- (\ref{eq:Lform})
and that magnetic fields are confined in the stellar interior,
the only non-trivial solutions are of dipole type.
Of course, once we relax the assumption of confined magnetic fields,
the existence of higher multipoles will be allowed. 

In the numerical analysis, it is convenient to integrate 
equation (\ref{eq:psi1}) from the stellar surface with
equation (\ref{eq:psibound}) as initial data.
By varying $\tilde L$, we look for a solution
 that behaves as regular as possible at the origin.
This procedure determines the eigenvalue $\tilde L$.
Next we integrate equation (\ref{eq:psi1}) from the origin
with the behavior~(\ref{eq:psi1reg})
and the fixed eigenvalue $\tilde L$.
Then the constant $\alpha_{1}$ is determined 
by demanding that the boundary condition at the
stellar surface, equation (\ref{eq:psibound}), is satisfied.

\section{Metric perturbation equations}
\label{sec:metric}

In this section, we derive the field equations for
the metric perturbation,
\beqa
\Delta g_{\mu \nu}=g_{\mu \nu}-\bar g_{\mu \nu}\,.
\eeqa
By linearizing the Einstein equations 
about the background metric $\bar g_{\mu \nu}$,
we have
\beqa
&&\left(\Delta g_{\mu \nu}{}_{;\alpha}{}^{;\alpha}
-\Delta g_{\mu \alpha}{}_{;\nu}{}^{;\alpha}
-\Delta g_{\nu \alpha}{}_{;\mu}{}^{;\alpha}
+\Delta g_{\alpha}{}^{\alpha}{}_{;\mu;\nu}\right)
\nonumber\\
&&\qquad
+\bar g_{\mu \nu} 
\left(\Delta g_{\alpha \beta}{}^{;\alpha;\beta}
-\Delta g_{\alpha}{}^{\alpha}{}_{;\beta}{}^{;\beta}
-\Delta g_{\alpha\beta} R^{\alpha \beta}\right)
+\Delta g_{\mu \nu} R=-16\pi \Delta T_{\mu \nu},
\label{eq:Ee}
\eeqa
where $\Delta T_{\mu \nu}$ is the perturbation of the energy momentum tensor,
and 
$R_{\mu\nu}$ and $R$ are the Ricci tensor and the Ricci scalar, respectively.
As shown below, the metric perturbation $\Delta g_{\mu \nu}$ is
 $O\left(\Psi^2\right)$.
Our method is similar to that developed by Hartle
(1967; Hartle \& Thorne 1968; Chandrasekhar \& Miller 1974)
for slowly rotating relativistic stars.
We obtain the field equations for the interior of the star 
in \S~\ref{sec:int},
and solve the exterior equations to give the junction conditions
at the stellar surface in \S~\ref{sec:ext}.

\subsection{Interior field equations}\label{sec:int}
The study of the metric perturbation in a spherical spacetime
was initiated by Regge \& Wheeler (1957)
and Zerilli (1970).
Because of the spherical symmetry of the background,
the perturbation equations are separable in the angular variables
by using ten tensor harmonics.
We can eliminate some components of the metric perturbation
with the aid of the gauge freedom.
In the gauge used by Regge \& Wheeler (1957),
the metric perturbation belonging to a given $\ell$ is of the form,
\beqa
\Delta g_{\mu \nu}=\left(
\begin{array}{cccc}
-2 e^{2\nu} H_{\ell} & I_{\ell} & 0 & 0 \\
I_{\ell} & \frac{2}{r} e^{4\lambda} M_{\ell} & 0 & 0 \\
0 & 0 & 2 r^2 K_{\ell} & 0 \\
0 & 0 & 0 & 2 r^2 \sin^2 \theta K_{\ell}
\end{array}
\right)
 P_{\ell}(\cos\theta)
%\nonumber\\
%&+&
+
\left(
\begin{array}{cccc}
0 & 0 & 0 & V_{\ell} \\
0 & 0 & 0 & W_{\ell} \\
0 & 0 & 0 & 0 \\
V_{\ell} & W_{\ell} & 0 & 0
\end{array}
\right)
%\times 
\sin\theta \frac{\partial}{\partial \theta} P_{\ell}(\cos\theta),
\label{eq:metricp}
\eeqa
where the first term is of parity $(-1)^{\ell}$ (even parity),
the second term is of parity $(-1)^{\ell+1}$ (odd parity)
and we may discard the $\varphi$-dependence of the harmonics
because of the axisymmetry.
Since there are fewer independent tensor harmonics for
 $\ell=0$ and $\ell=1$, 
we may further simplify the metric by adopting a gauge
in which
$I_{0}=0$, $K_{0}=0$, $V_{0}=0$, $W_{0}=0$, $K_{1}=0$
 and $W_{1}=0$ (Zerilli 1970).

The energy momentum tensor perturbation $\Delta T_{\mu \nu}$
is obtained by taking the perturbation of equation (\ref{eq:emten}).
The components of the magnetic field to $O(\Psi)$ are given
from equations (\ref{eq:Bu}), (\ref{eq:uinv}) -- (\ref{eq:tuperp}),
(\ref{eq:order1}), (\ref{eq:order2}) and (\ref{eq:wind2nd})
as
\beqa
B^{0}&=&O\left(\Psi^3\right),
\label{eq:B0}
\\
B^{1}&=&-\frac{2}{r^2} e^{-\lambda} \psi_{1} \cos\theta,
\label{eq:B1}
\\
B^{2}&=&\frac{1}{r^2} e^{-\lambda} \psi_{1}' \sin\theta,
\label{eq:B2}
\\
B^{3}&=&\frac{\tilde L}{r^2} e^{-\nu} \psi_{1}\,.
\label{eq:B3}
\eeqa
Since the energy momentum tensor is quadratic in $B^\mu$, 
we need $\Delta T_{\mu\nu}$ to the accuracy of $O\left(\Psi^2\right)$.
To this order, from equations (\ref{eq:uinv}) -- (\ref{eq:tuperp}),
(\ref{eq:order1}), (\ref{eq:order2}) and (\ref{eq:wind2nd}),
the fluid flow is given by
\beqa
u^{0}&=&u_{\eta}=\frac{1}{\sqrt{-g_{00}}},
\label{eq:u0}
\\
u^{1}&=&\frac{2}{\tilde C \rho r^2} e^{-\lambda-\nu}
\left(\psi_{1}\right)^2 \sin^2\theta \cos\theta,
\label{eq:u1}
\\
u^{2}&=&-\frac{1}{\tilde C \rho r^2} e^{-\lambda-\nu}
\psi_{1} \psi_{1}' \sin^3\theta,
\label{eq:u2}
\\
u^{3}&=&\Omega u_{\eta}+u_{\xi}
=\Omega e^{-\nu}-\frac{\tilde L}{\tilde C \rho r^2} e^{-2\nu}
\left(\psi_{1}\right)^2 \sin^2\theta.
\label{eq:u3}
\eeqa
Expanding the perturbation of the enthalpy 
in equation (\ref{eq:defmu}) in the form,
\beqa
\frac{\Delta \mu}{\mu}
=\sum_{\ell=0}^{\infty}
 \left[\Delta \ln \mu(r)\right]_{\ell} P_{\ell}(\cos\theta),
\eeqa
we obtain from equations (\ref{eq:wind2nd}) and (\ref{eq:metricp}) 
the equations
\beqa
(\Delta \ln \mu)_{0}&=&-H_{0}-\frac{2}{3}\frac{D_{1}}{D_{0}}\psi_{1}+
\frac{D_{2}}{D_{0}},
\label{eq:lnmu0}
\\
(\Delta \ln \mu)_{2}&=&-H_{2}+\frac{2}{3}\frac{D_{1}}{D_{0}}\psi_{1},
\label{eq:lnmu2}
\eeqa
and $(\Delta \ln \mu)_{\ell}=-H_{\ell}$ for $\ell \ne 0,2$. 
{}From the first law of thermodynamics, equation (\ref{eq:1law}),
with equation (\ref{eq:defmu}),
we have the relations $(\Delta p)_{\ell}=\rho (\Delta \mu)_{\ell}$ and 
$(\Delta \epsilon)_{\ell}=(p/\rho^2) (\Delta \rho)_{\ell}$,
where $(\Delta p)_{\ell}$, $(\Delta \epsilon)_{\ell}$ 
and $(\Delta \rho)_{\ell}$ are the $\ell$-th pole perturbations
of the pressure, the internal energy and 
the rest mass density, respectively.
For the polytropic equation of state $p=K \rho^{1+1/n}$,
we have $(\Delta p)_{\ell}=(1+1/n) (p/\rho) (\Delta \rho)_{\ell}$.

{}From the Einstein equations~(\ref{eq:Ee}) with 
$\Delta g_{\mu\nu}$ given by equation (\ref{eq:metricp}),
we obtain the  following independent equations which have nonzero
source terms, 
\beqa
&&M_0'=4\pi r^2 \mu^2 \rho \frac{d\rho}{dp} (\Delta \ln \mu)_0+
\frac{1}{3}\left(\tilde L^2 e^{-2\nu}
+\frac{2}{r^2}\right) \left(\psi_{1}\right)^2
+\frac{1}{3}e^{-2\lambda} \left(\psi_{1}'\right)^2,
\quad \left(G_{00}\right)
\label{eq:M0}
\\
&&H_{0}'=-\left[(\Delta \ln \mu)_{0}\right]'
-\frac{2}{3\mu} e^{-\nu} D_{1} \psi_{1}'
\nonumber\\
&&\quad 
=\left(\frac{1}{r^2}+8\pi p\right) e^{4\lambda} M_{0}
+4\pi r e^{2\lambda} \rho \mu (\Delta \ln \mu)_{0}
%\nonumber\\
%&&\quad
+\frac{1}{3 r} e^{2\lambda}\left(\tilde L^2
e^{-2\nu}-\frac{2}{r^2}\right) \left(\psi_{1}\right)^2
+\frac{1}{3 r} \left(\psi_{1}'\right)^2,
\quad \left(G_{11}\right)
\label{eq:H0}
\\
&&I_{1}=-\frac{32\pi}{5}\frac{\mu}{\tilde C}
e^{\lambda} \left(\psi_{1}\right)^2,
\quad \left(G_{01}\right)
\label{eq:I1}
\\
&&I_{3}=\frac{16\pi}{15}\frac{\mu}{\tilde C}
e^{\lambda} \left(\psi_{1}\right)^2,
\quad \left(G_{01}\right)
\label{eq:I3}
\\
&&V_{1}''-\left(\nu'+\lambda'\right) V_{1}'
-\frac{2}{r}\left(\nu'+\lambda'+\frac{1}{r}\right) V_{1}
=\frac{64\pi}{5} \frac{\tilde L}{\tilde C} \mu
e^{2\lambda-\nu} \left(\psi_{1}\right)^2
-16\pi \rho\mu r^2 e^{2\lambda} \Omega,
\quad \left(G_{03}\right)
\label{eq:V1}
\\
&&V_{3}''-\left(\nu'+\lambda'\right) V_{3}'
-\frac{2}{r}\left(\nu'+\lambda'+\frac{1}{r}
+\frac{5}{r} e^{2\lambda}\right) V_{3}
=-\frac{32\pi}{15} \frac{\tilde L}{\tilde C} \mu
e^{2\lambda-\nu} \left(\psi_{1}\right)^2,
\quad \left(G_{03}\right)
\label{eq:V3}
\\
&&W_{2}=-\frac{2}{3} \tilde L e^{\lambda-\nu}
\left(\psi_{1}\right)^2,
\quad \left(G_{13}\right)
\label{eq:W2}
\\
&&H_{2}+\frac{1}{r} e^{2\lambda} M_{2}=
\frac{2}{3} e^{-2\lambda} \left(\psi_{1}'\right)^2
-\frac{2}{3} \tilde L^2 e^{-2\nu}
\left(\psi_{1}\right)^2,
\quad \left(G_{22}-G_{33}/\sin^2\theta\right)
\label{eq:M2}
\\
&&\left(\nu'+\frac{1}{r}\right) K_{2}'
-\frac{2}{r^2} e^{2\lambda} K_2
+\frac{1}{r} H_2'
-\frac{3}{r^2} e^{2\lambda} H_2
-\frac{1}{r}\left(\frac{1}{r^2}+8\pi p\right) e^{4\lambda} M_2
\nonumber\\
&&\qquad =4\pi e^{2\lambda} \rho \mu (\Delta \ln \mu)_2
-\frac{1}{3r^2} e^{2\lambda}
\left(\tilde L^2 e^{-2\nu}+\frac{4}{r^2}\right)
\left(\psi_{1}\right)^2
-\frac{1}{3r^2} \left(\psi_{1}'\right)^2,
\quad \left(G_{11}\right)
\label{eq:H2}
\\
&&K_{2}'+H_{2}'-\left(\frac{1}{r}-\nu'\right) H_{2}
-\frac{1}{r}\left(\frac{1}{r}+\nu'\right) e^{2\lambda} M_{2}
=\frac{4}{3 r^2} \psi_{1} \psi_{1}',
\quad \left(G_{12}\right)
\label{eq:K2}
\eeqa
where we have indicated the component of the field equation
in the parentheses at the end of each equation
from which it is derived.
The remaining components of the Einstein equations that are not
listed in the above are either trivially satisfied or
not independent from the above set because of
the Bianchi identities (Zerilli 1970).
%\beqa
%Y_2=H_2+K_2
%\eeqa
%
%\beqa
%&&Y_2'=-2\nu' H_2+ \frac{2}{3}\left(\frac{1}{r}+\nu'\right)
%\left[e^{-2\lambda} \left(\psi_{1}'\right)^2
%-\tilde L^2 e^{-2\nu} \left(\psi_{1}\right)^2\right]
%+\frac{4}{3r^2}\psi_{1} \psi_{1}'
%\\
%&&H_2'=-\frac{2}{r^2 \nu'} e^{2\lambda} Y_2
%+\left[-2\nu'+\frac{1}{\nu'} e^{2\lambda} 
%\left(4\pi \rho \mu-\frac{2 m}{r^3}\right)\right] H_2
%\nonumber\\
%&&\quad
%+\frac{1}{3}\left(2\nu' e^{-2\lambda}+\frac{1}{r^2\nu'}\right)
%\left(\psi_{1}'\right)^2
%+\frac{1}{3}\left(-2\nu' e^{-2\lambda}+\frac{1}{r^2\nu'}\right)
%\tilde L^2 e^{2(\lambda-\nu)} \left(\psi_{1}\right)^2
%\nonumber\\
%&&\quad
%+\frac{4}{3r^2}\left(1+\frac{1}{r \nu'}\right)
%\psi_{1} \psi_{1}'
%+\frac{4}{3r^4 \nu'} e^{2\lambda} \left(\psi_{1}\right)^2
%-\frac{8\pi}{3\nu'} \rho e^{2\lambda-\nu} D_{1} \psi_{1}
%\eeqa
To avoid the cancellation of significant digits
in the numerical calculation, it is convenient to define
\beqa
Y_2=H_2+K_2-\frac{1}{6} e^{-2\lambda} \left(\psi_{1}'\right)^2
-\frac{2}{3r} e^{-2\lambda} \psi_{1} \psi_{1}'
-\frac{2}{3r^2} \left(\psi_{1}\right)^2,
\label{eq:defY2}
\eeqa
and solve for $Y_2$ instead of $K_2$.
With equations (\ref{eq:lnmu2}) and (\ref{eq:M2}),
we can rewrite equations (\ref{eq:H2}) and (\ref{eq:K2}) as
\beqa
&&Y_2'+2\nu' H_2=
\nu' e^{-2\lambda} \left(\psi_{1}'\right)^2
+\frac{1}{3}\left[\tilde L^2 e^{-2\nu}
+\frac{2}{r}\left(\nu'+\lambda'\right) e^{-2\lambda}
-\frac{4 m}{r^3}
\right] \psi_{1} \psi_{1}'
\nonumber\\
&&\quad
-\frac{2}{3} \nu' \tilde L^2 e^{-2\nu} \left(\psi_{1}\right)^2
-\frac{4\pi}{3} r \rho e^{-\nu} D_{1} \left(r \psi_{1}'+2\psi_{1}\right),
\label{eq:Y2}
\\
&&H_2'+\frac{2}{r^2 \nu'} e^{2\lambda} Y_2
+\left[2\nu'-\frac{1}{\nu'} e^{2\lambda} 
\left(4\pi \rho \mu-\frac{2 m}{r^3}\right)\right] H_2
\nonumber\\
&&\quad
=\frac{2}{3}\nu' e^{-2\lambda} \left(\psi_{1}'\right)^2
+\frac{4}{3r^2} \psi_{1} \psi_{1}'
+\frac{1}{3}\left(-2\nu'+\frac{1}{r^2\nu'} e^{2\lambda}\right)
\tilde L^2 e^{-2\nu} \left(\psi_{1}\right)^2
-\frac{8\pi}{3\nu'} \rho e^{2\lambda-\nu} D_{1} \psi_{1}.
\label{eq:H22}
\eeqa
We have six differential equations (\ref{eq:M0}), (\ref{eq:H0}),
(\ref{eq:V1}), (\ref{eq:V3}), (\ref{eq:Y2}) and (\ref{eq:H22})
for six unknown functions; $M_0$, $(\Delta \ln \mu)_0$,
$V_1$, $V_3$, $Y_2$ and $H_2$.
A practical method to solve these equations is discussed in the next section.
The functions $H_0$, $I_1$, $I_3$, $W_2$, $M_2$, $K_2$ and $(\Delta \ln \mu)_2$
are determined algebraically from equations (\ref{eq:lnmu0}), 
(\ref{eq:I1}), (\ref{eq:I3}), (\ref{eq:W2}), (\ref{eq:M2}), 
(\ref{eq:defY2}) and (\ref{eq:lnmu2}), respectively.
%The other components are obviously zero.

\subsection{Exterior solutions and boundary conditions}\label{sec:ext}
The equations derived in the previous subsection
apply to the interior of the star.
Outside the star, in the vacuum,
we have $\psi_{1}=0$, $m(r)=M_{*}$, $e^{2\nu}=1-2M_{*}/r$ and 
$\Delta \ln \mu=0$ and the resulting equations can be solved explicitly.
The solutions for which the metric $g_{\mu \nu}$ is asymptotically flat
are obtained as
\beqa
M_0&=&\Delta M_{*},
\label{eq:M0e}
\\
H_0&=&-\frac{\Delta M_{*}}{r-2M_{*}},
\label{eq:H0e}
\\
I_1&=&I_3=0,
\\
V_1&=&\frac{2 \Delta J}{r},
\label{eq:V1e}
\\
V_3&=&\frac{\Delta V}{8M_{*}^3}\, V_{f}\left(\frac{r}{2 M_{*}}\right),
\label{eq:V3e}
\\
W_2&=&0,
\\
H_2&=&\frac{5 \Delta Q}{8 M_{*}^3}\, Q_{2}^{2}\left(\frac{r}{M_{*}}-1\right),
\label{eq:H2e}
\\
Y_2&=&-\frac{5 \Delta Q}{8 M_{*}^3}
\frac{2M_{*}}{[r(r-2M_{*})]^{1/2}}\, Q_{2}^{1}\left(\frac{r}{M_{*}}-1\right),
\label{eq:Y2e}
\\
M_2&=&-(r-2M_{*})H_2,
\label{eq:M2e}
\eeqa
where $\Delta M_{*}$, $\Delta J$, $\Delta V$ and $\Delta Q$ are constants,
$V_f$ is the function defined by
\beqa
V_{f}(z)=\frac{7}{2}
\left[\frac{1}{z}+5+30z-210z^2+180z^3+60z^2\left(3z^2-5z+2\right)
\ln \left(1-\frac{1}{z}\right)\right],
\eeqa
and $Q_{n}^{m}$ are the associated Legendre functions of the second kind,
\beqa
Q_{2}^{2}(z)&=&\frac{z\left(5-3z^2\right)}{z^2-1}
+\frac{3}{2}\left(z^2-1\right)\ln \frac{z+1}{z-1},
\\
Q_{2}^{1}(z)&=&\frac{2-3z^2}{\left(z^2-1\right)^{1/2}}
+\frac{3}{2} z \left(z^2-1\right)^{1/2} \ln \frac{z+1}{z-1}.
\eeqa
In the limit $z\to \infty$, we have $V_{f}(z)\to {1}/{z^3}$,
$Q_{2}^{2}(z)\to {8}/{5 z^3}$
and $Q_{2}^{1}(z)\to -{2}/{5 z^3}$.
We may identify $\Delta M_{*}$ and $\Delta J$
with the mass shift and the angular momentum of the star, respectively,
while $\Delta Q$ and $\Delta V$
with the mass quadrupole and current hexapole moments,
respectively (Thorne 1980),
apart from numerical factors.

Let us now explain our strategy to solve six unknown functions $M_{0}$,
$(\Delta \ln \mu)_{0}$, $V_{1}$, $V_{3}$, $Y_{2}$ and $H_2$.
First, to obtain $M_{0}$ and $(\Delta \ln \mu)_{0}$,
we need to specify what quantity is to be fixed 
when we add the perturbation.
Although we may fix the central density
(Hartle 1967; Hartle \& Thorne 1968; Chandrasekhar \& Miller 1974)
 or the total mass, here we adopt the baryon mass as the fixed parameter.
{}From equation (\ref{eq:bacon}), the total baryon mass is given by
(Hartle 1967)
\beqa
M_{B}=\int d^3 x \sqrt{-g} \rho u^{0}.
\eeqa
Then, from equations (\ref{eq:metricp}) and (\ref{eq:u0}),
the perturbation of the total baryon mass up to $O(\Psi^2)$
is given by
\beqa
\Delta M_{B}=\int_{0}^{R_{*}} dr 4\pi r^2 e^{\lambda}
\left[\frac{1}{r} e^{2\lambda} \rho M_{0}+(\Delta \rho)_0\right].
\eeqa
This may be written in the differential form,
\beqa
\Delta M_{B}'(r)
=4\pi r^2 e^{\lambda}\left[\frac{1}{r} e^{2\lambda} \rho M_{0}
+\rho \mu \frac{d\rho}{dp} (\Delta \ln \mu)_{0}\right],
\label{eq:dbmass}
\eeqa
where $\Delta M_B(r)$ is the baryon mass within radius $r$,
and we have used the equality
$(\Delta\rho)_0=\rho\mu (d\rho/dp) (\Delta\ln\mu)_0$
for an isentropic fluid.
To fix the baryon mass, we solve this equation with
the boundary conditions $\Delta M_{B}(0)=0$ and $\Delta M_{B}(R_{*})=0$,
together with equations (\ref{eq:M0}) for $M_0$ and
 (\ref{eq:H0}) for $(\Delta \ln \mu)_{0}$.
Near the origin these functions have the behaviors,
\beqa
M_0 &\to& \frac{1}{3} r^3\left[4\pi \mu_{c}^2 \rho_{c} \frac{d\rho_{c}}{dp_{c}}
(\Delta \ln \mu)_{0c}+2 \alpha_{1}^2\right],
\\
(\Delta \ln \mu)_{0} &\to& (\Delta \ln \mu)_{0c},
\\
\Delta M_{B} &\to& \frac{4\pi}{3} r^3 \mu_{c} \rho_{c} \frac{d\rho_{c}}{dp_{c}}
(\Delta \ln \mu)_{0c},
\eeqa
where $\alpha_{1}$ is the constant
introduced in equation (\ref{eq:psi1reg}), i.e.,
$\psi_1\to r^2 \alpha_{1}$, and
the subscript 'c' means the value at the center of the star.
With these initial behaviors,
equations (\ref{eq:M0}), (\ref{eq:H0}) and (\ref{eq:dbmass})
can be integrated from the center to the stellar surface.
We first determine the central value $(\Delta \ln \mu)_{0c}$
so as to satisfy the boundary condition $\Delta M_{B}(R_{*})=0$.
Then the value of $M_0$ at $r=R_{*}$ gives the mass shift $\Delta M_{*}$
from equation (\ref{eq:M0e}),
and the solution for $(\Delta \ln \mu)_{0}$, 
together with equations (\ref{eq:lnmu0}) 
and (\ref{eq:H0e}), determines the constant $D_{2}$.

Next, to obtain $V_{1}$, it is convenient to express it
as the sum of a particular solution and a homogeneous solution
\beqa
V_{1}=V_{1}^{(P)}+c_{1} V_{1}^{(H)},
\label{eq:V1pV1h}
\eeqa
where $c_{1}$ is a constant and the homogeneous solution $V_{1}^{(H)}$
satisfies equation (\ref{eq:V1}) with the right-hand side being zero.
Near the origin, we have
\beqa
V_{1}^{(P)} &\to& r^2, 
\quad
V_{1}^{(H)} \to r^2.
\label{eq:V1r0}
\eeqa
Then, $V_{1}^{(P)}$ and $V_{1}^{(H)}$ can be obtained
by integrating equation (\ref{eq:V1}) from the center with the
 initial behavior~(\ref{eq:V1r0}) to the stellar surface.
The constant $c_{1}$ in equation (\ref{eq:V1pV1h})
and the angular momentum $\Delta J$ in equation (\ref{eq:V1e})
can be determined by requiring the continuities
of the value $V_{1}$ and its derivative $V_{1}'$
with the exterior solution in equation (\ref{eq:V1e}).

A similar method can be applied to equation (\ref{eq:V3}) for $V_3$.
In this case, the behavior near the origin is
\beqa
V_{3}^{(P)} &\to& r^4, 
\quad
V_{3}^{(H)} \to r^4.
\eeqa
Then the solution determines the current hexapole moment $\Delta V$
from equation (\ref{eq:V3e}).

Finally, to obtain $Y_{2}$ and $H_{2}$,
we again express them as the sum of a particular solution
and a homogeneous solution
\beqa
Y_{2}=Y_{2}^{(P)}+c_{2} Y_{2}^{(H)},
\quad
H_{2}=H_{2}^{(P)}+c_{2} H_{2}^{(H)},
\label{eq:YH2}
\eeqa
where $c_{2}$ is a constant and
the homogeneous solutions $Y_{2}^{(H)}$ and $H_{2}^{(H)}$
satisfy equations (\ref{eq:Y2}) and (\ref{eq:H22}), respectively,
with the right-hand side being zero.
Near the origin, they behave as
\beqa
Y_{2}^{(P)} &\to& \frac{r^4}{6}
\left(
%-4\pi\left(\rho_{c}+\rho_{c} \epsilon_{c}+3 p_{c}\right) H_{c}^{(P)}
\tilde L^2 e^{-2\nu_c} \alpha_{1}^2
-8\pi \rho_{c} e^{-\nu_c} D_{1} \alpha_{1}\right),
\quad
H_{2}^{(P)} \to 
\frac{8}{3}r^2 \alpha_{1}^2,
%r^2\left[H_{c}^{(P)}
%+\frac{8}{3}\left(\alpha_{1}\right)^2\right],
\label{eq:YH2pr0}
\\
Y_{2}^{(H)} &\to& -\frac{2\pi}{3} r^4 
\left(\rho_{c}+\rho_{c} \epsilon_{c}+3p_{c}\right),
\quad
H_{2}^{(H)} \to r^2.
\label{eq:YH2hr0}
\eeqa
These functions can be calculated
by integrating equations (\ref{eq:Y2}) and (\ref{eq:H22})
from the center with the initial behaviors~(\ref{eq:YH2pr0}) and
(\ref{eq:YH2hr0}) to the stellar surface.
The constant $c_{2}$ in equation (\ref{eq:YH2})
and the mass quadrupole moment $\Delta Q$ in equations (\ref{eq:H2e})
and (\ref{eq:Y2e}) can be determined by requiring the continuities
of $Y_{2}$ and $H_{2}$
with the exterior solutions~(\ref{eq:H2e}) and (\ref{eq:Y2e}).

%With these boundary conditions,
%the magnetized star has the same central density as the non-magnetized star.
%$V_{1}$ and $V_{3}$ should be finite at the origin.
%$Y_{2}$ and $H_{2}$ vanish at the origin.
%\beqa
%M_{0} &\to& \frac{2}{3} r^3 \left(\alpha_{1}\right)^2
%\\
%\delta (\ln \mu)_{0} &\to& 
%-\frac{2}{3} r^2 \left(\alpha_{1}\right)^2
%-\frac{2}{3} \frac{E_{1}}{E_{0}} r^2 \alpha_{1}
%=-\frac{2}{3} r^2 \left(\alpha_{1}\right)^2
%-\frac{2}{3\mu} e^{-\nu} E_{1} r^2 \alpha_{1}
%\\
%\eeqa

\section{Numerical results}\label{sec:results}
In this section we present the results of our numerical
integration of the linearized GS equation~(\ref{eq:psi1})
and the metric perturbation equations in \S~\ref{sec:metric}.
We adopt the fifth-order Runge-Kutta method with the adaptive stepsize control
(Press et al. 1992) and use it iteratively when necessary.
We adopt the polytropic equation of state
\beqa
p=K\rho^{1+1/n},
\eeqa
with the polytropic index $n=1$.
We consider the non-rotating case $\Omega=0$
to concentrate on the magnetic effects.

\subsection{Magnetic field and fluid flow structure}
The GS equation~(\ref{eq:psi1}) for $\psi_{1}$
is the basic equation to obtain the fluid flow 
and the electromagnetic field. The boundary condition~(\ref{eq:psibound})
is satisfied for discrete eigenvalues $\tilde L$.
The first few roots of the (dimensionless) eigenvalue
$R_{*} \tilde L$ are given in Table~\ref{tab:eigen}
for several relativistic factors $M_{*}/R_{*}$.
Although $\tilde L$ can take both signs,
we assume $\tilde L>0$ in the following.

\begin{figure}
\plottwo{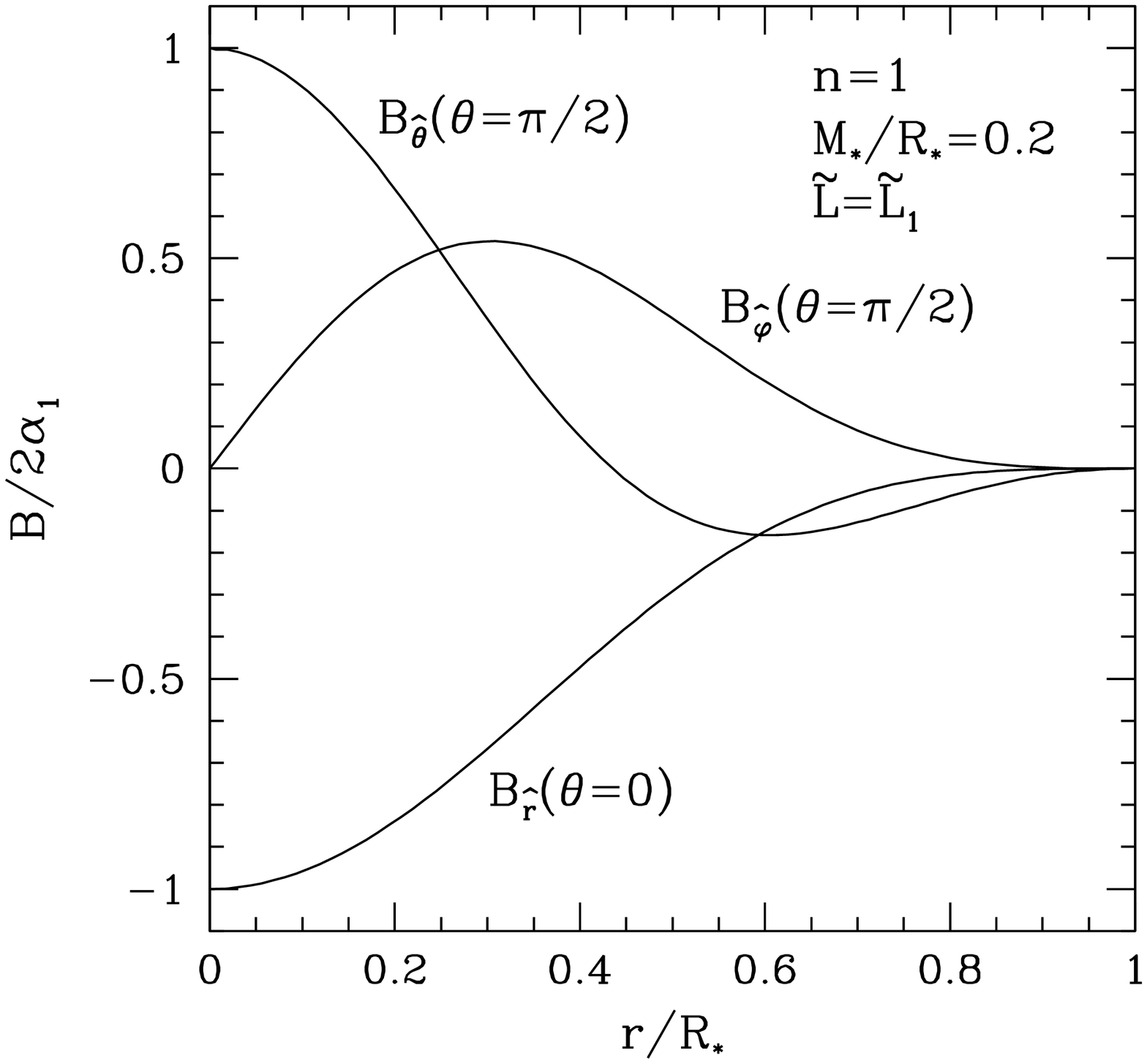}{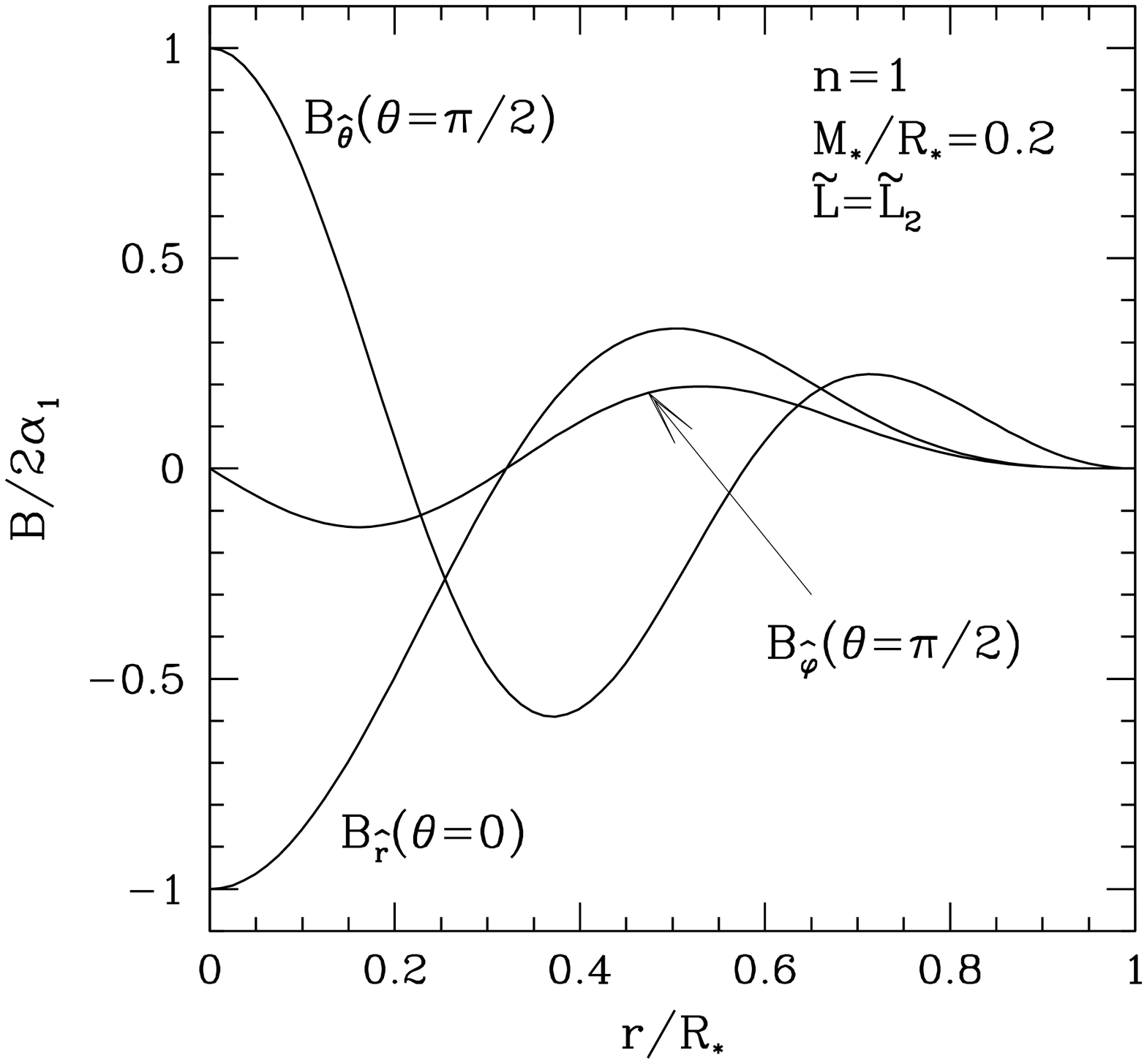}
\caption{\label{fig:mf12}
The components $B_{\hat r}(\theta=0)$,
$B_{\hat \theta}(\theta=\pi/2)$
and $B_{\hat \varphi}(\theta=\pi/2)$
of the magnetic field in the orthonormal basis
in equations (\ref{eq:B(r)}) -- (\ref{eq:B(v)}) 
normalized by the central magnetic field $2\alpha_{1}$
are shown as functions of radius $r/R_{*}$.
We adopt the polytropic index $n=1$,
the relativistic factor $M_{*}/R_{*}=0.2$.
The left panel shows the case of the eigenvalue $\tilde L=\tilde L_{1}$ 
and the right panel the case of $\tilde L=\tilde L_{2}$
given in Table~\ref{tab:eigen}.
}
\end{figure}

\begin{figure}
\plottwo{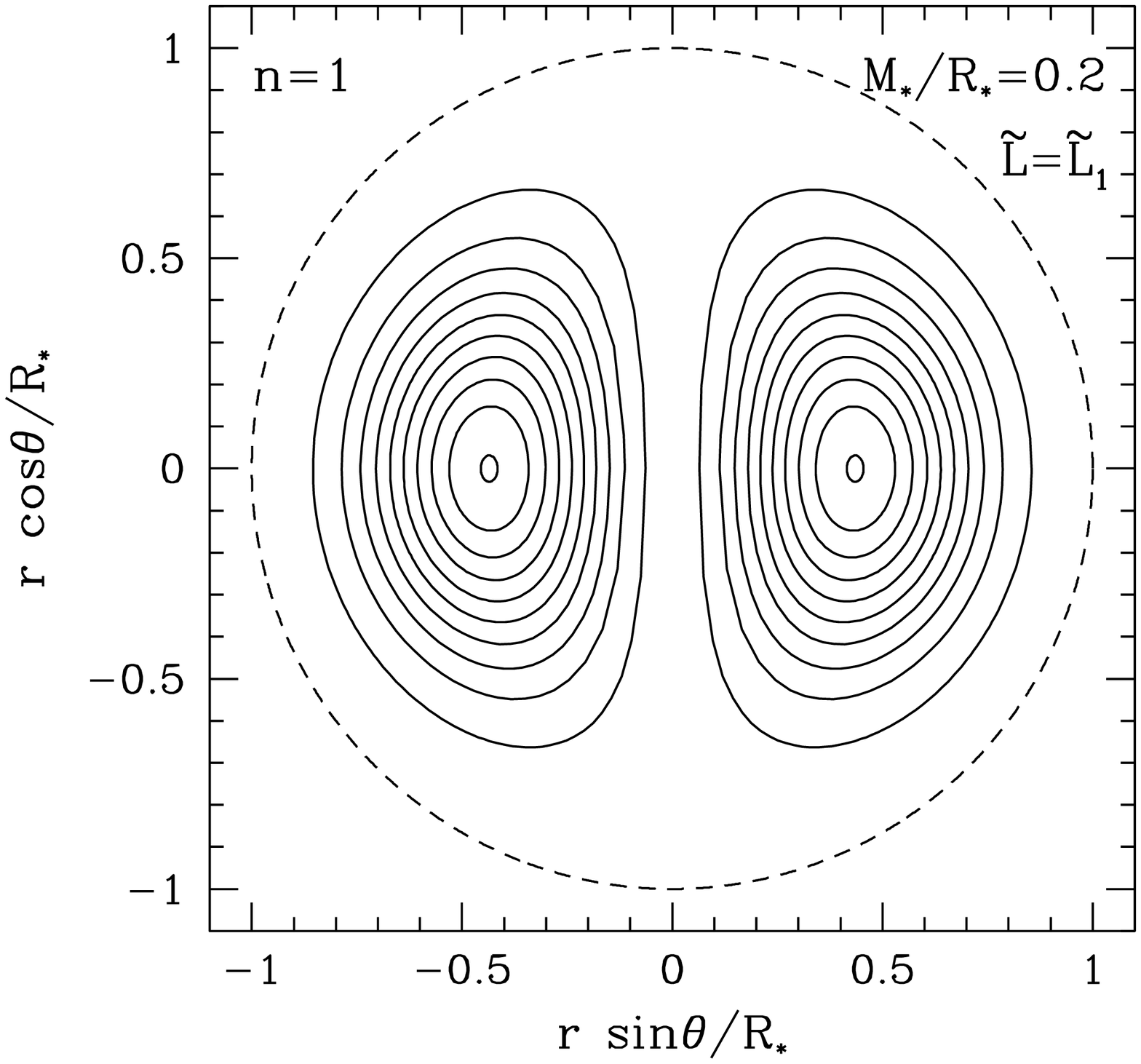}{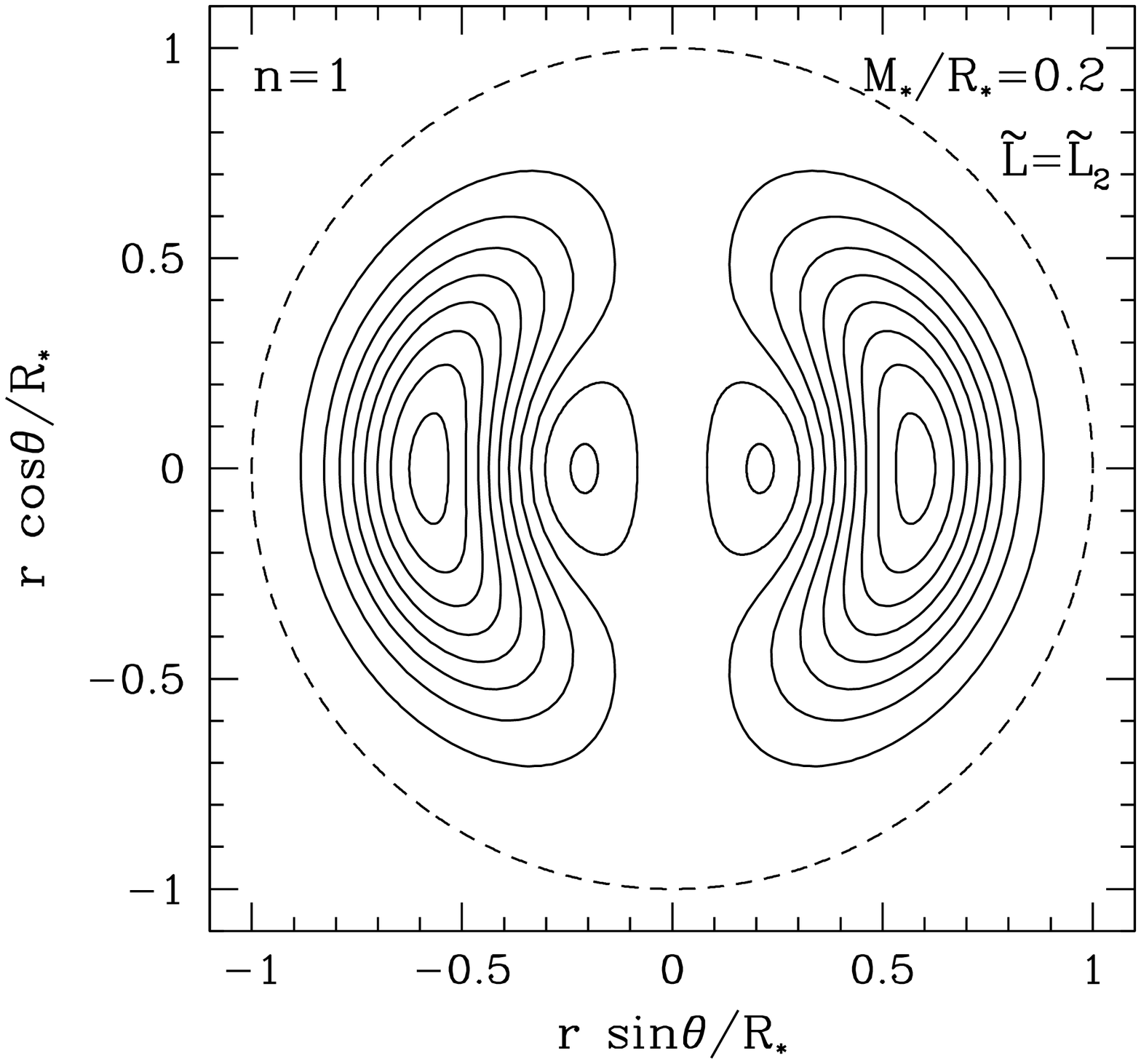}
\caption{\label{fig:mtop12} 
The magnetic field lines projected on the meridional plane,
i.e., the $\Psi=$const. lines in the $(r,\theta)$-plane.
The dashed line is the stellar surface.
We adopt the polytropic index $n=1$,
the relativistic factor $M_{*}/R_{*}=0.2$.
The left panel shows the case of the eigenvalue $\tilde L=\tilde L_{1}$
and the right panel the case of $\tilde L=\tilde L_{2}$
given in Table~\ref{tab:eigen}.
}
\end{figure}

\begin{figure}
\plotone{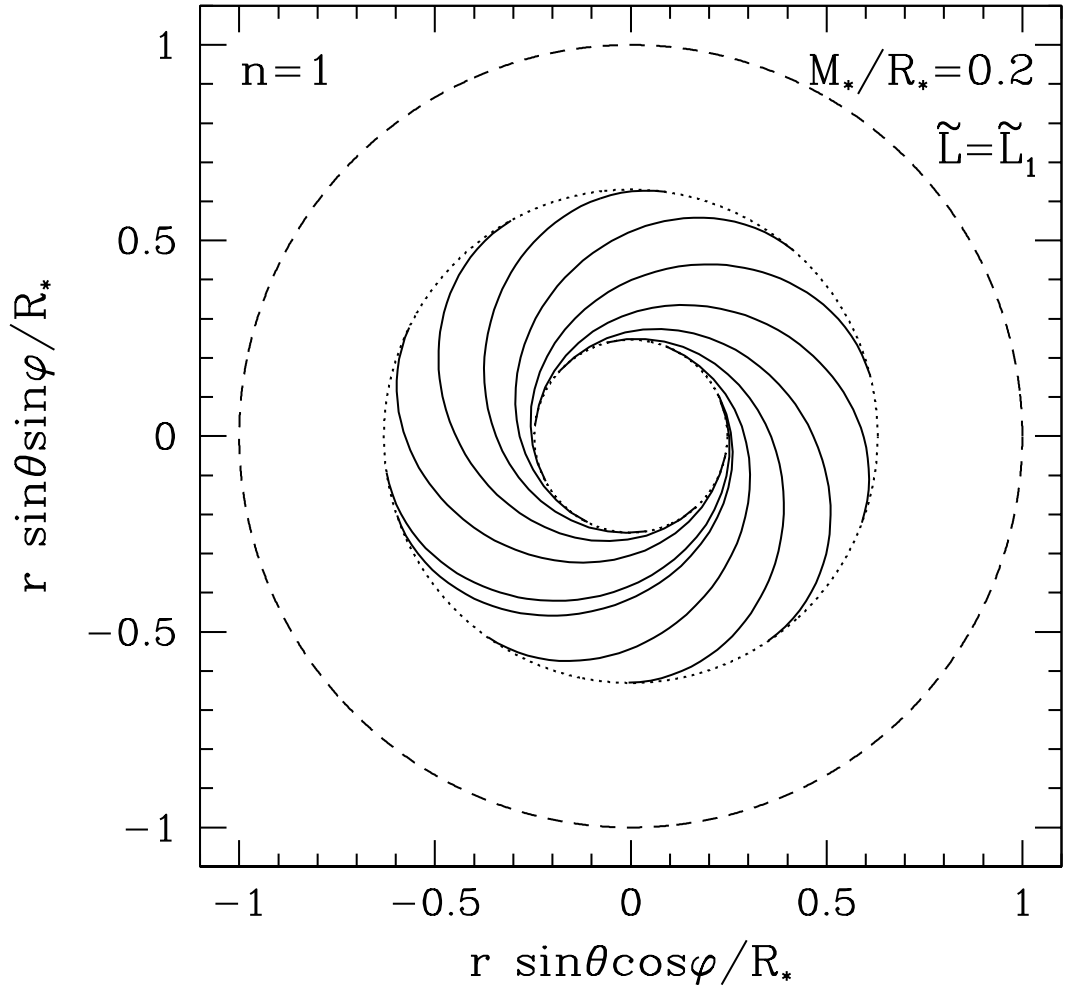}
\caption{\label{fig:wind}
A truncated piece of the magnetic field line on a certain flux surface
($\Psi=$const. surface) projected onto the equatorial plane 
($\theta=\pi/2$ plane). Only the part of it
in the upper hemisphere ($\theta<\pi/2$) is shown. 
The dashed line is the stellar surface.
The dotted lines are the cross section of the flux surface 
with the equatorial plane.
We adopt the polytropic index $n=1$,
the relativistic factor $M_{*}/R_{*}=0.2$,
and the eigenvalue $\tilde L=\tilde L_{1}$ in Table~\ref{tab:eigen}.
}
\end{figure}

The components of the magnetic field $B^\mu$
are given by equations (\ref{eq:B0}) -- (\ref{eq:B3}).
Although the magnetic field is defined in the fluid frame
in equation (\ref{eq:B}), the fluid 4-velocity $u^\mu$ there
may be identified with the unit vector $n^\mu$ normal to 
the $t=$const. hypersurface to the accuracy of $O(\Psi)$.
Therefore, it is convenient to express the magnetic
field in terms of the orthonormal tetrad components
in the background metric~(\ref{eq:backmetric}).
They are given by
\beqa
B_{\hat r}&=&-\frac{2}{r^2} \psi_{1} \cos\theta,
\label{eq:B(r)}
\\
B_{\hat \theta}&=&\frac{1}{r} e^{-\lambda} \psi_{1}' \sin\theta,
\label{eq:B(t)}
\\
B_{\hat \varphi}&=&\frac{\tilde L}{r} e^{-\nu} \psi_{1} \sin\theta.
\label{eq:B(v)}
\eeqa
In Figures \ref{fig:mf12}, we show $B_{\hat r}(\theta=0)$, 
$B_{\hat \theta}(\theta=\pi/2)$ and $B_{\hat \varphi}(\theta=\pi/2)$
normalized by $2 \alpha_{1}$ as a function of radius $r$
for the smallest two eigenvalues of $\tilde L$.
Note that the magnetic field strength at the center of the star
is given by $B_{c}^{2}=(2 \alpha_{1})^2$.
{}From these figures, we can see that the magnetic structure is
more complicated for a larger eigenvalue $\tilde L$.
To be precise, the number of nodes at which the $r$-derivative 
of the magnetic field vanishes increases as $\tilde L$ becomes large.
This is also illustrated in Figures \ref{fig:mtop12},
in which the magnetic field lines projected on the meridional plane,
i.e., the $\Psi=$const. lines in the $(r,\theta)$-plane, are shown.
In Figure \ref{fig:wind}, we show the projection of 
a truncated piece of a magnetic field line onto the equatorial plane.
Only the part of it in the upper hemisphere ($\theta<\pi/2$) is shown. 
As we can see from the figure, the flux surface forms a torus and
the magnetic field line winds around the torus.

\begin{figure}
\plottwo{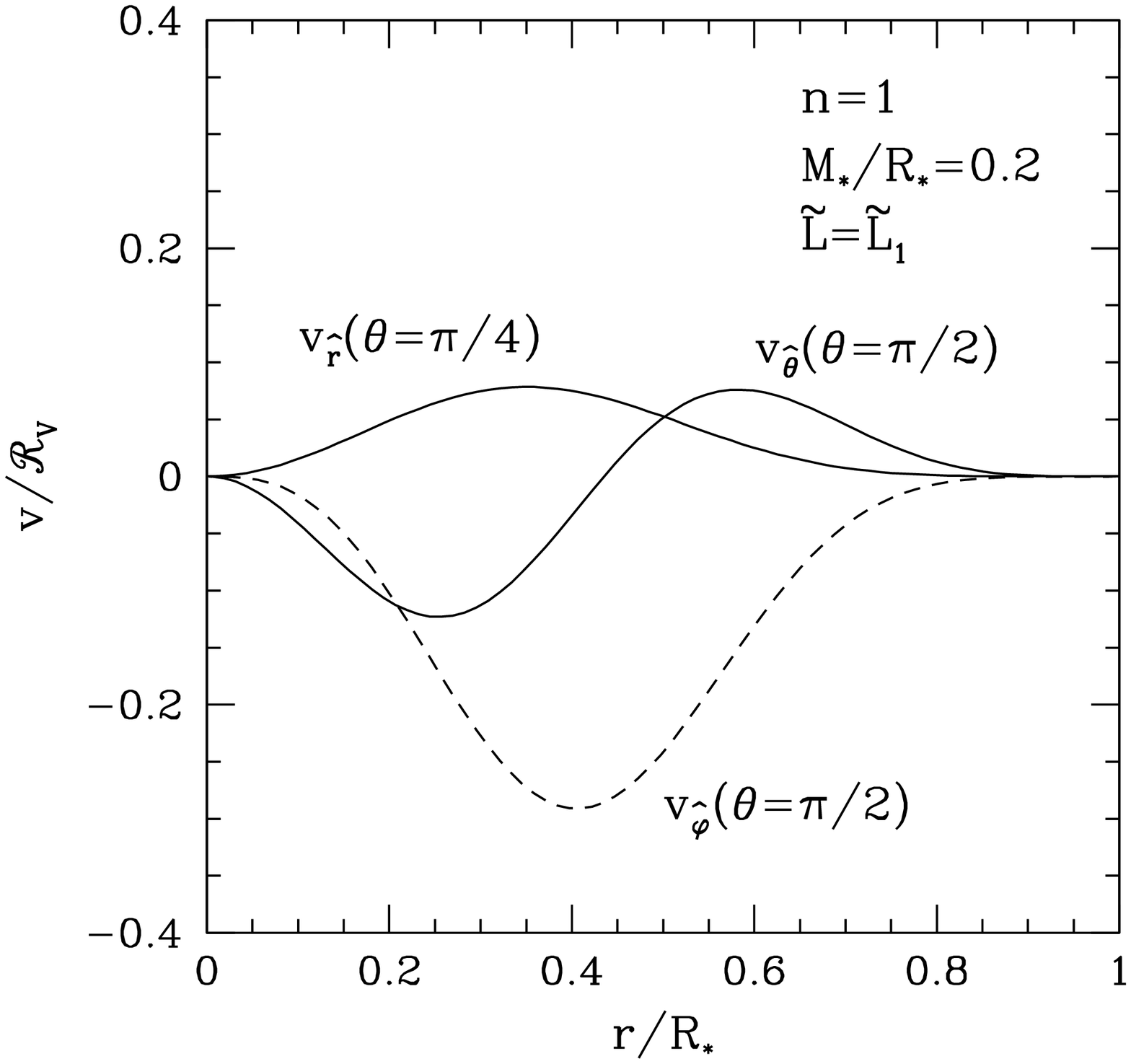}{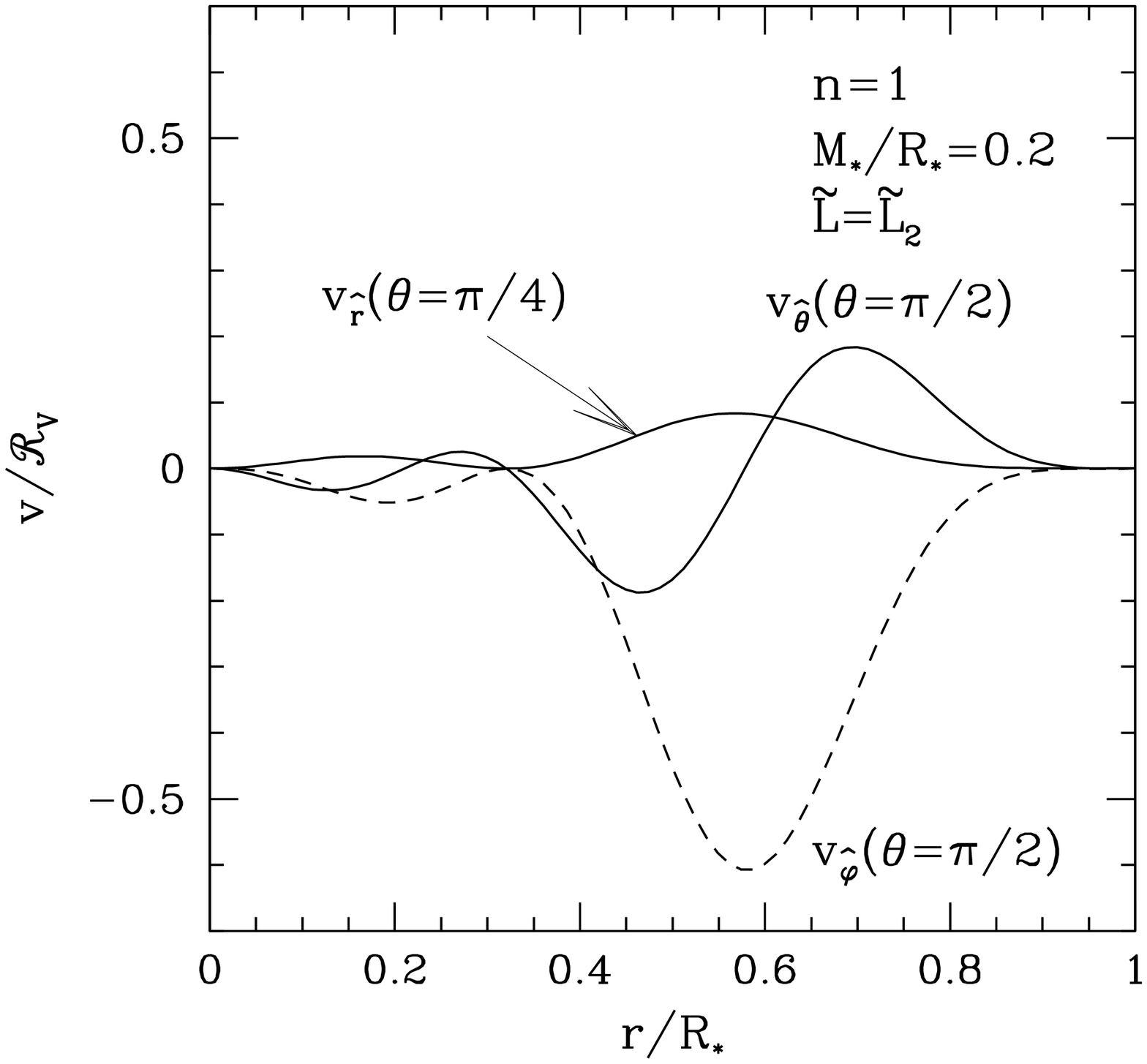}
\caption{\label{fig:velo12} 
The components $v_{\hat r}(\theta=\pi/4)$,
$v_{\hat \theta}(\theta=\pi/2)$
and $v_{\hat \varphi}(\theta=\pi/2)$
of the fluid velocity in the orthonormal basis
in equations (\ref{eq:v(r)}) -- (\ref{eq:v(v)}),
normalized by the dimensionless variable ${\cal R}_{V}$ in equation 
(\ref{eq:Rv}),
as functions of radius $r/R_{*}$.
We adopt the polytropic index $n=1$,
the relativistic factor $M_{*}/R_{*}=0.2$.
The left panel shows the case of the
eigenvalue $\tilde L=\tilde L_{1}$ 
and the right panel the case of $\tilde L=\tilde L_{2}$
given in Table~\ref{tab:eigen}.
}
\end{figure}

\begin{figure}
\plotone{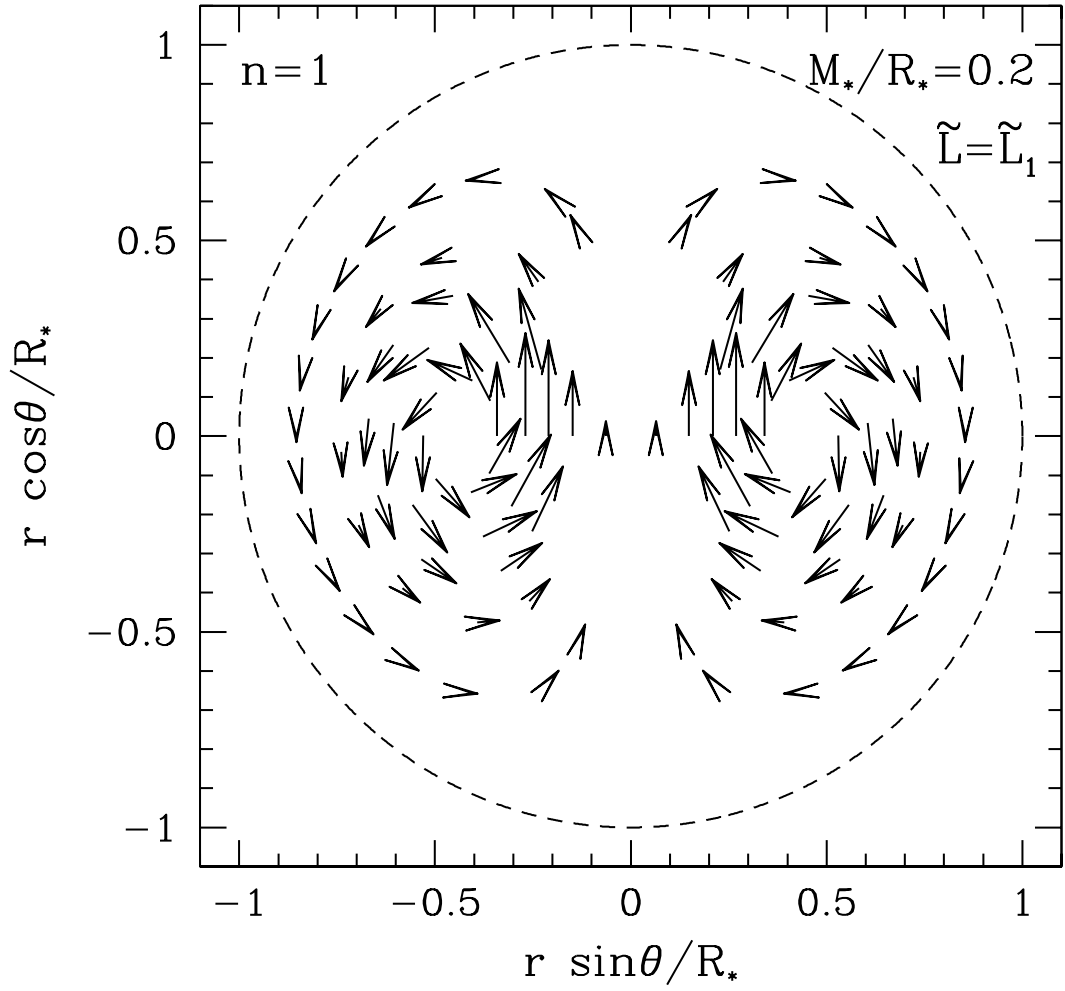}
\caption{\label{fig:flow1}
The velocity field in the meridional plane ($\varphi=$const. plane).
The lengths of the vectors are proportional to the fluid velocity
in the orthonormal basis in equations (\ref{eq:v(r)}) -- (\ref{eq:v(v)}).
The dashed line is the stellar surface.
We adopt the polytropic index $n=1$,
the relativistic factor $M_{*}/R_{*}=0.2$,
and the eigenvalue $\tilde L=\tilde L_{1}$ in Table~\ref{tab:eigen}.
}
\end{figure}

The fluid flow is determined from equations (\ref{eq:u0}) -- (\ref{eq:u3}).
The orthonormal tetrad components of the fluid velocities 
in the background metric~(\ref{eq:backmetric})
are given by
\beqa
v_{\hat r}&=&\frac{2}{\tilde C \rho r^2} e^{-\nu}
\left(\psi_{1}\right)^2 \sin^2\theta \cos\theta,
\label{eq:v(r)}
\\
v_{\hat \theta}&=&-\frac{1}{\tilde C \rho r} e^{-\lambda-\nu}
\psi_{1} \psi_{1}' \sin^3\theta,
\label{eq:v(t)}
\\
v_{\hat \varphi}&=&\Omega e^{-\nu} r\sin\theta
-\frac{\tilde L}{\tilde C \rho r} e^{-2\nu}
\left(\psi_{1}\right)^2 \sin^3\theta,
\label{eq:v(v)}
\eeqa
where we set $\Omega=0$ in the present analysis.
In Figures \ref{fig:velo12},
we show $v_{\hat r}(\theta=\pi/4)$, $v_{\hat \theta}(\theta=\pi/2)$
and $v_{\hat \varphi}(\theta=\pi/2)$ as functions of radius $r$
for the smallest two eigenvalues of $\tilde L$,
normalized by the dimensionless quantity,
\beqa
{\cal R}_{V}:=\frac{M_{*} R_{*}}{\tilde C}
        \frac{R_{*}^4 \alpha_{1}^2}{M_{*}^2}\,.
\label{eq:Rv}
\eeqa
As in the case of the magnetic field,
the velocity field is more complicated for a larger eigenvalue.
In Figure \ref{fig:flow1}, we plot the velocity field on the meridional plane
($\varphi=$const. plane).
The flow line resides on the constant flux surface.

In equation (\ref{eq:Rv}), the dimensionless quantity,
\beqa
{\cal R}_{M}:=\frac{R_{*}^4 \alpha_{1}^2}{M_{*}^2}
= \frac{3}{2}\frac{(B_{c}^2/8\pi)(4\pi R_{*}^3/3)}{M_{*}^2/R_{*}},
\label{eq:Rm}
\eeqa
describes the ratio of the magnetic energy to the gravitational energy,
apart from a numerical factor.
The quantity $\tilde C$ controls the magnitude of the meridional flow.
As mentioned in \S~\ref{sec:1order},
our perturbation method is valid as long as
\beqa
|\tilde C| \gg |R_{*}^3 \alpha_{1}| = O(\Psi)\,.
\eeqa
The limit $|\tilde C| \to \infty$ corresponds to 
the limit of zero meridional flow.

\subsection{Mass shift and magnetic helicity}
By integrating the metric perturbation equations in \S~\ref{sec:metric}
according to the prescription of \S~\ref{sec:ext},
we can calculate the change in the central density $(\Delta \rho)_{0c}$,
the mass shift $\Delta M_{*}$, the angular momentum $\Delta J$,
the mass quadrupole moment $\Delta Q$ 
and the current hexapole moment $\Delta V$.
In Table~\ref{tab:integ}, we list the numerical results 
for several configurations.
Each value is normalized appropriately as indicated in the first column.
Here the dimensionless variable ${\cal R}_{V}$ defined 
in equation (\ref{eq:Rv}) measures the magnitude of the meridional flow
velocity and ${\cal R}_{M}$ defined in equation (\ref{eq:Rm})
measures the ratio of the magnetic energy to the gravitational energy.
We also introduce the dimensionless magnetic helicity ${\cal H}_{M}$ 
as defined in equation (\ref{eq:mh}) below.
The numbers given in Table~\ref{tab:integ} can be easily converted
to physical units once we specify the mass $M_{*}$ and
the dimensionless variables ${\cal R}_{V}$ (or $\tilde C$) and
 ${\cal R}_{M}$ (or $\alpha_{1}$).

The mass shift $\Delta M_{*}/M_{*} {\cal H}_{M}$
increases for larger eigenvalue $\tilde L$.
This means that the configuration with the smallest eigenvalue 
$\tilde L=\tilde L_{1}$ is the lowest energy state 
for a fixed baryon mass $M_{B}$ and total
magnetic helicity ${\cal H}_{M}$.
The magnetic helicity is conserved in a perfect
conducting plasma (Woltjer 1958),
and it describes topological properties
such as links, twists and kinks of magnetic field lines
(Moffatt 1969; Berger \& Field 1984).
When the topology of the field lines changes,
the conductivity should be finite, which implies
non-conservation of the magnetic helicity.
However, it was conjectured that the total magnetic helicity 
is still approximately conserved on reconnection or 
relaxation time scales (Taylor 1974; Berger 1984).
This hypothesis has been successful in laboratory plasmas (Taylor 1986).
It is uncertain whether or not the total magnetic helicity is nearly
conserved in the neutron star.
In any case, we may speculate that the configuration with
the smallest eigenvalue is preferred if the free energy could be
released by the rearrangement
of the magnetic field (see also Ioka 2001).

In the covariant language, the 4-current of magnetic helicity
is defined (Carter 1978; Bekenstein 1987)
by
\beqa
H^{\mu}=\frac{1}{2}\epsilon^{\mu \nu \alpha \beta}
 A_{\nu} F_{\alpha \beta}.
\eeqa
This is conserved in the ideal MHD,
\beqa
H^{\mu}{}_{;\mu}=0.
\eeqa
Thus the total magnetic helicity is defined by
$\int d^3 x \sqrt{-g} H^{0}$.
Although the 4-current of magnetic helicity $H^{\mu}$
is not gauge-invariant locally, 
the total magnetic helicity is gauge-invariant 
for confined magnetic fields.
{}From equations (\ref{eq:F12}), (\ref{eq:u0}) and (\ref{eq:u3}) we have
\beqa
F_{12}=\tilde L e^{\lambda-\nu} \psi_{1} \sin\theta\,.
\eeqa
Hence, using the gauge freedom to take $A_{2}=0$, we find
\beqa
A_{1}=\tilde L e^{\lambda-\nu} \psi_{1} \cos\theta.
\eeqa
Therefore, we define the (dimensionless) total magnetic helicity by
\beqa
{\cal H}_{M}:=\frac{1}{M_{*}^2}\int d^3x \sqrt{-g} H^{0}
=\frac{1}{M_{*}^2} \int_{0}^{R_{*}} dr 
\frac{16 \pi}{3} \tilde L e^{\lambda-\nu} \left(\psi_{1}\right)^2.
\label{eq:mh}
\eeqa

\subsection{Spherical and quadrupolar deformations}
The magnetic stress distorts a star.
The isobaric ($p$=const.) surface that lies at a radius $r_0$ 
in the background configuration
is displaced in the perturbed configuration to a radius
\beqa
r=r_{0}+(\Delta r)_{0}(r_{0})+(\Delta r)_{2}(r_{0}) P_{2}(\cos\theta),
\eeqa
where, from equation (\ref{eq:nu'}) and 
$\left(\Delta p\right)_{\ell}=\rho\left(\Delta \mu\right)_{\ell}$,
we have 
\beqa
(\Delta r)_{\ell}=\frac{1}{\nu'} (\Delta \ln \mu)_{\ell}
\quad (\ell=0,2).
\eeqa
The spherical deformation is characterized by the
monopole displacement $(\Delta r)_{0}$. 
However, to describe the quadrupolar 
deformation in a geometrically invariant manner,
the coordinate radius $r$ is not quite appropriate.
Instead of $r$, the circumferential radius 
$r_{*}^2=r^2[1+2 K_{2} P_{2}(\cos\theta)]$ is more appropriate
(Hartle \& Thorne 1968).
Then the isobaric surface is expressed as
\beqa
r_{*}(r_0,\theta)=r_{0}+(\Delta r)_{0}(r_{0})
+\left[(\Delta r)_{2}(r_{0})+r_0 K_{2}(r_{0})\right] P_{2}(\cos\theta).
\eeqa
We describe the quadrupolar deformation 
by the ellipticity defined by
\beqa
e_{*}(r)=\frac{r_{*}(r,\pi/2)-r_{*}(r,0)}{r}
=-\frac{3}{2}\left[\frac{\left(\Delta \ln\mu\right)_{2}}{r \nu'}
+K_{2}\right].
\eeqa

In Table~\ref{tab:integ}, we also list $(\Delta r)_{0}$ and $e_{*}$
at the stellar surface $r=R_{*}$ for various configurations.
We can see that the stars are more deformed
for larger eigenvalues of $\tilde L$,
for the baryon mass and total magnetic helicity fixed.
This is intuitively consistent with the results in the previous
subsection that the energy (the mass shift) is higher for
larger $\tilde L$
for constant baryon mass and total magnetic helicity.
We also find that the configurations are prolate rather than
oblate since $e_{*}<0$.
This is in contrast with the case of
stars with only poloidal fields which are oblate rather 
than prolate (Bocquet et al. 1995; Konno, Obata, \& Kojima 1999).
This difference is due to the fact that
the magnetic stress of a toroidal field tends to make a star prolate,
working like a rubber belt tightening up
the equator of the star (Ostriker \& Gunn 1969; Ioka 2001).
Our model is the first example that demonstrates the validity of 
this picture in relativistic stars.

%%%%%%%%%%%%%%%%%%%%%%%%%%%%%%%%%%%%%
\subsection{Frame dragging}
In general relativity, inertial frames are dragged 
not only by the fluid flow but also by the magnetic field.
The dragging of inertial frames manifests itself 
in the off-diagonal components of the metric~(\ref{eq:metricp}).
In addition to the $t\varphi$-component ($V_{\ell}$) which
is familiar in the case of rotating stars and Kerr black holes,
there also exist the $tr$-component ($I_{\ell}$)
and the $r\varphi$-component ($W_{\ell}$) in the Regge-Wheeler gauge.
These components originate from the meridional flow
and the toroidal magnetic field,
which are incorporated into the relativistic stellar model
for the first time in this paper.

The 4-velocity of an observer whose world line is orthogonal to
the $t=$const. hypersurface $\Sigma_{t}$ is given by $n^{\mu}$ 
in equation (\ref{eq:defn}).
To the lowest order in $\Psi$, the coordinate angular velocity 
and the coordinate radial velocity of such observers are given by
\beqa
\frac{d\varphi}{dt}&=&\frac{n^{3}}{n^{0}}=
N^{\varphi}=\frac{V_{1}}{r^2}+\frac{V_{3}}{r^2} \frac{3}{2}
\left(5\cos^2\theta-1\right),
\\
\frac{dr}{dt}&=&\frac{n^{1}}{n^{0}}=N^{1}
=-e^{-2\lambda} I_{1} \cos\theta
-e^{-2\lambda} I_{3} \frac{1}{2}\left(5\cos^3\theta-3\cos \theta\right),
\label{eq:drdt}
\eeqa
from equations (\ref{eq:ncom}), (\ref{eq:metric}) and (\ref{eq:metricp}).

\begin{figure}
\plottwo{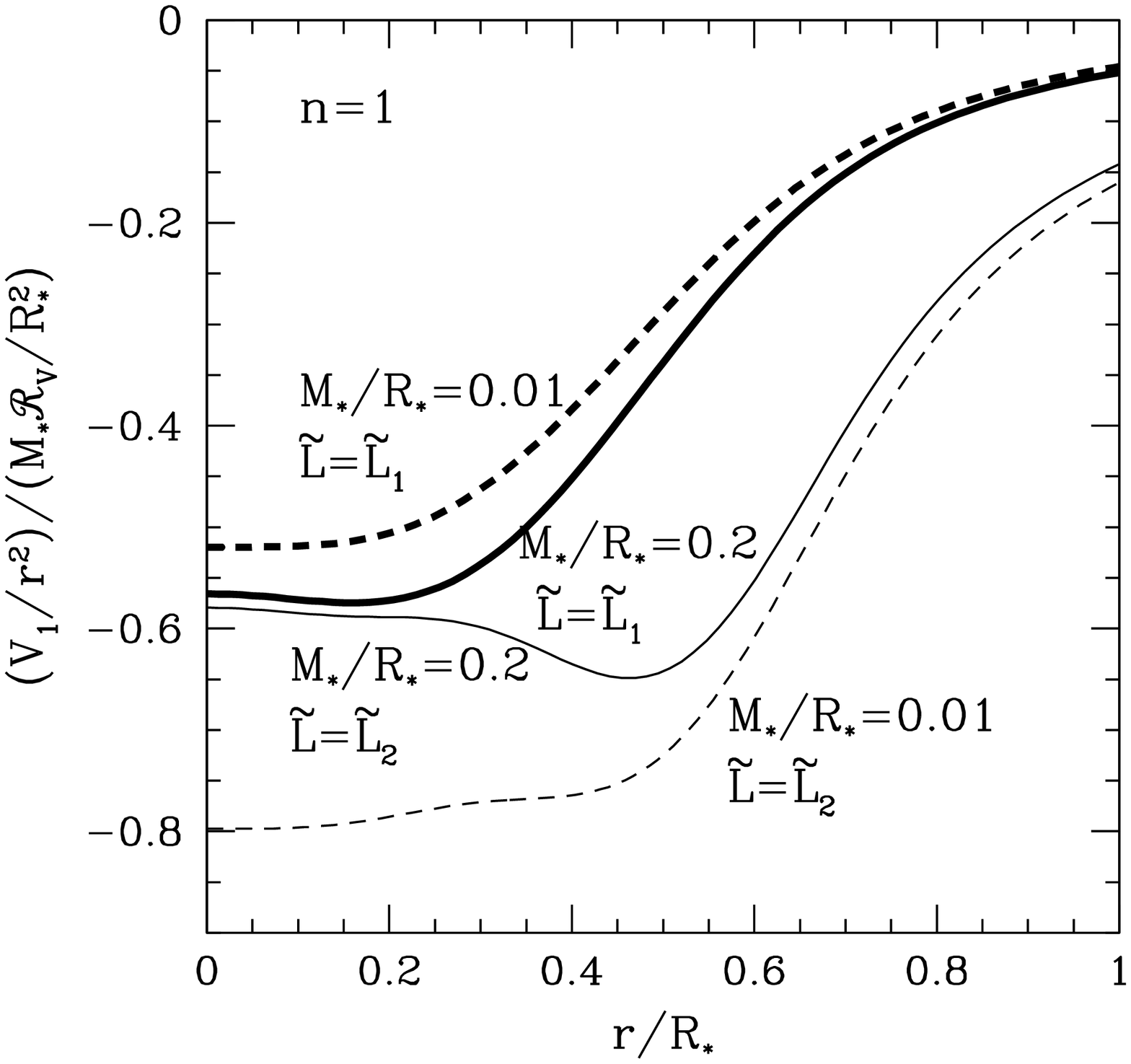}{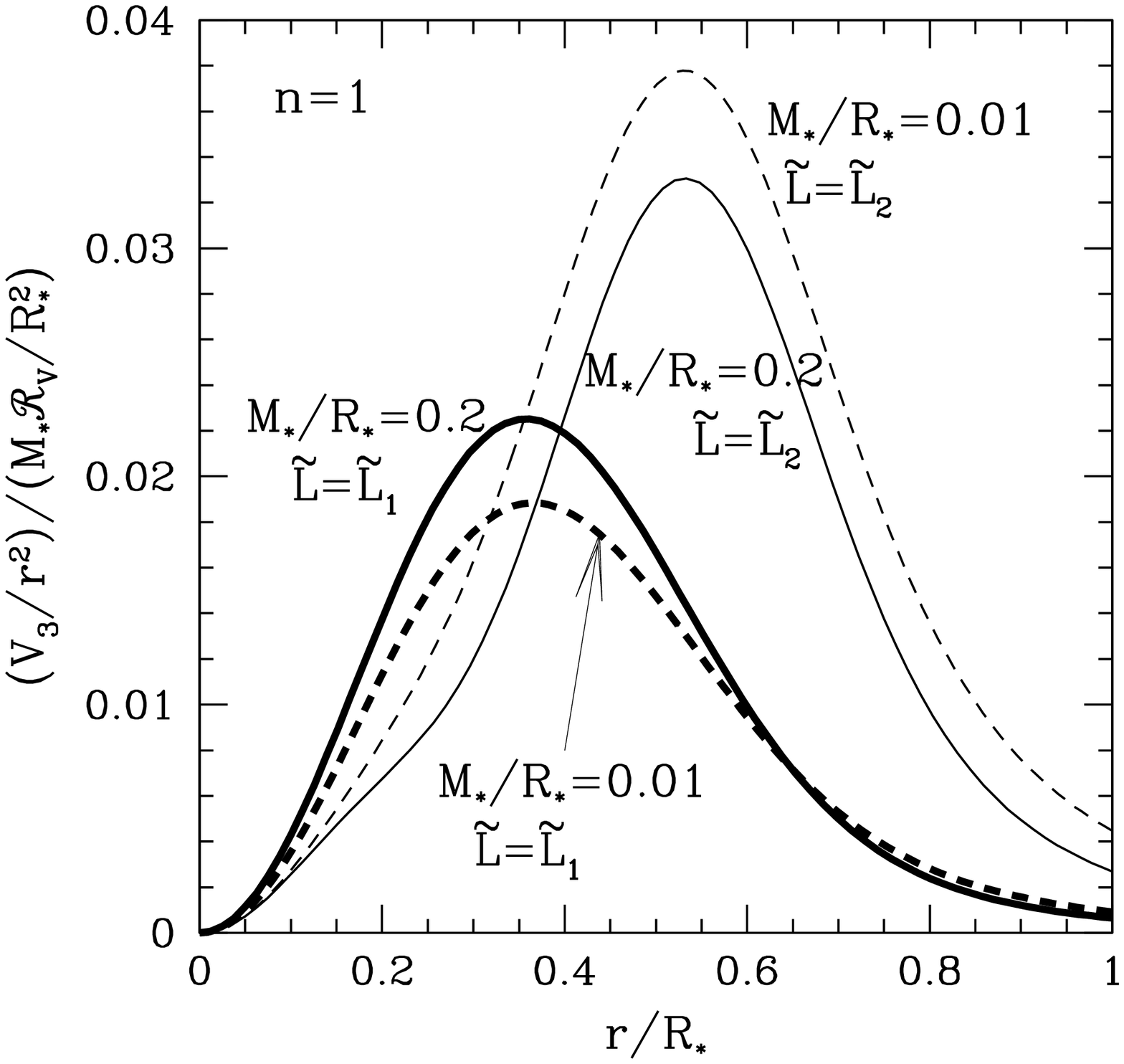}
\caption{\label{fig:v13}
The $t\varphi$-component of the dipole ($\ell=1$) metric perturbation
(left panel) and that of the hexapole ($\ell=3$) metric perturbation
(right panel) as functions of radius,
namely, $V_{1}/r^2$ normalized by $M_{*}{\cal R}_{V}/R_{*}^2$
and $V_{3}/r^2$ normalized by $M_{*}{\cal R}_{V}/R_{*}^2$,
respectively.
The solid and dashed lines are for $M_{*}/R_{*}=0.2$ (relativistic)
and $M_{*}/R_{*}=0.01$ (Newtonian), respectively,
while the bold and thin lines for $\tilde L=\tilde L_{1}$
and $\tilde L=\tilde L_{2}$ in Table~\ref{tab:eigen},
respectively.
We adopt the polytropic index $n=1$.
These metric components describe the standard frame dragging 
due to the angular momentum.
}
\end{figure}

In Figures \ref{fig:v13},
we show the $t\varphi$-component of the metric perturbation
that describes the coordinate angular velocity,
for $M_{*}/R_{*}=0.2$ (relativistic case) and $M_{*}/R_{*}=0.01$
(Newtonian case). 
The left panel shows the dipole ($\ell=1$) component of the
(normalized) coordinate angular velocity
$(V_{1}/r^2)/(M_{*}{\cal R}_{V}/R_{*}^2)$
and the right panel shows the hexapole ($\ell=3$) component
$(V_{3}/r^2)/(M_{*}{\cal R}_{V}/R_{*}^2)$.
From these figures, we can see that 
the normalized coordinate angular velocity 
does not depend on the relativistic factor $M_{*}/R_{*}$ so much.
Thus, if the velocity of the meridional flow is fixed, i.e., 
${\cal R}_{V}=$const.,
the coordinate angular velocity is nearly proportional 
to $(M_{*}/R_{*})^2$ for fixed mass.
Note that the $t\varphi$-component ($V_{\ell}$) is gauge-invariant
for stationary perturbations (see equation (D3) in Zerilli 1970).

%%%%%%%%%%%%%%%%%%%%%%%%%%%%%%%%%%%%%%%%%

\begin{figure}
\plottwo{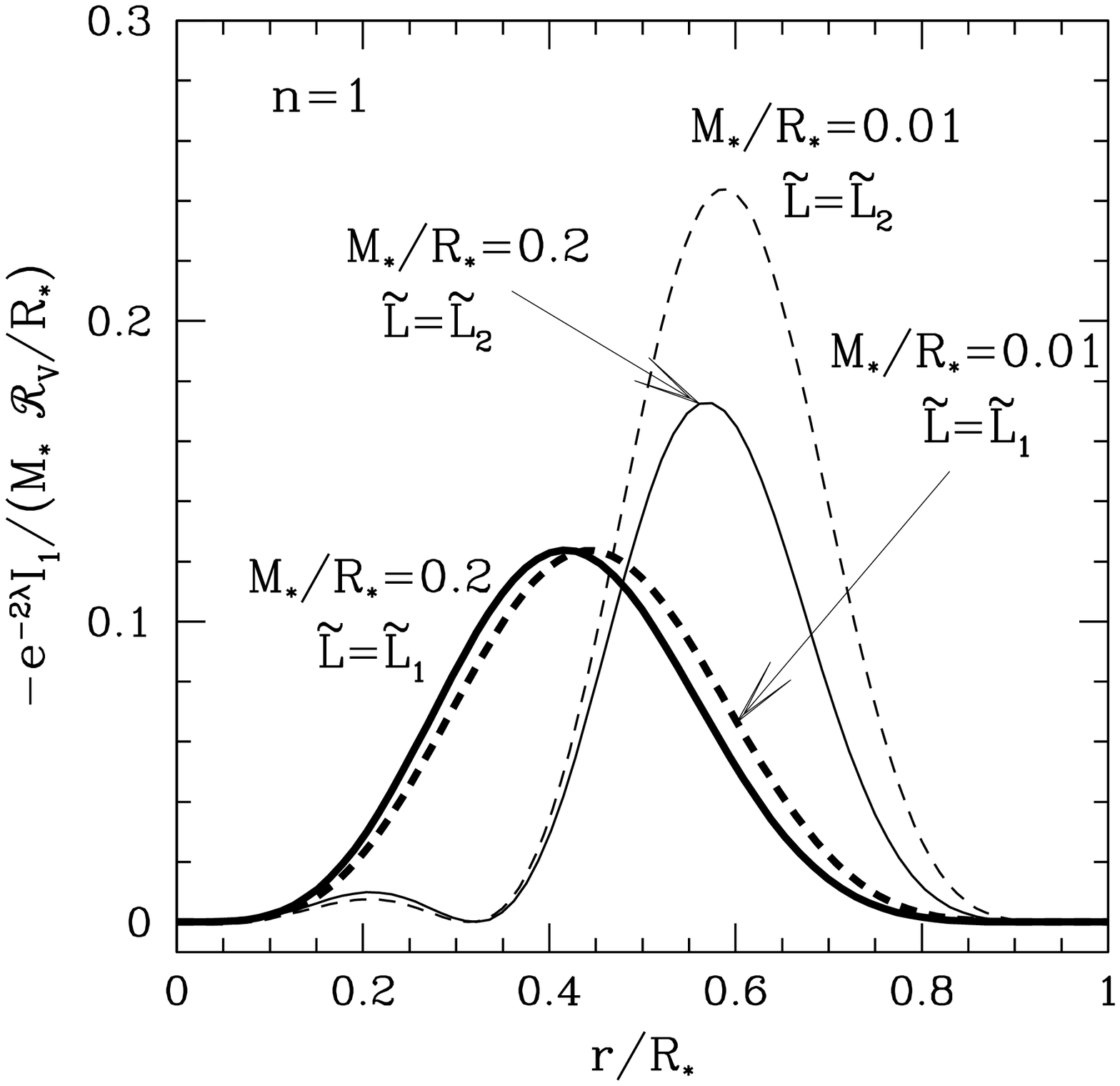}{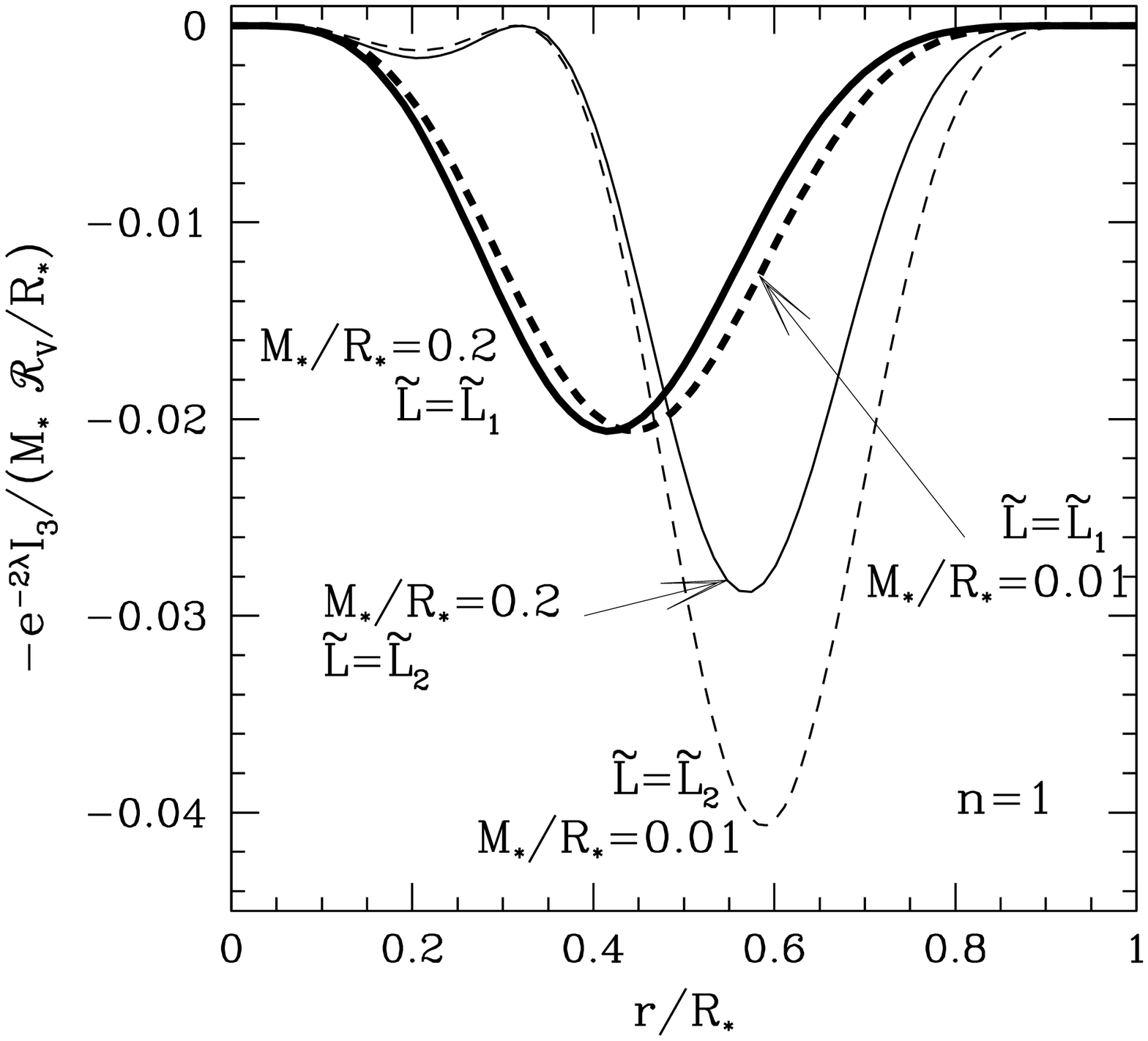}
\caption{\label{fig:i13} 
The $tr$-component of the dipole ($\ell=1$) metric perturbation
(left panel) and that of the hexapole ($\ell=3$) metric perturbation
(right panel) as functions of radius,
namely,
$-e^{-2\lambda} I_{1}$ normalized by $M_{*}{\cal R}_{V}/R_{*}$
and 
$-e^{-2\lambda} I_{3}$ normalized by $M_{*}{\cal R}_{V}/R_{*}$,
respectively.
The solid and dashed lines are for $M_{*}/R_{*}=0.2$ (relativistic)
and $M_{*}/R_{*}=0.01$ (Newtonian), respectively, while
the bold and thin lines for $\tilde L=\tilde L_{1}$ 
and $\tilde L=\tilde L_{2}$ in Table~\ref{tab:eigen}, respectively.
We adopt the polytropic index $n=1$.
These metric components describe a new type of frame dragging
due to meridional flow.
}
\end{figure}

In Figures \ref{fig:i13},
we show the dipole and hexapole components of the
(normalized) coordinate radial velocity in equation (\ref{eq:drdt}),
$-e^{-2\lambda} I_{1}/(M_{*}{\cal R}_{V}/R_{*})$
and $-e^{-2\lambda} I_{3}/(M_{*}{\cal R}_{V}/R_{*})$, respectively.
From these figures, and from the left panel of Figures \ref{fig:velo12}
where we see that the fluid velocity $v$ is about $v \sim 0.1 {\cal R}_{V}$,
as an order of magnitude estimate,
the coordinate radial velocity is found to be $\sim (M_{*}/R_{*}) v$.
It is interesting to note that the presence of the
components $I_{1}$ and $I_{3}$ violates the reflection symmetry
about the equatorial plane,
i.e., $I_{1}$ and $I_{3}$ are of parity $(-1)$ in equation (\ref{eq:metricp}).
Note that the $tr$-component ($I_{\ell}$) is gauge-specific
by itself (see equation (D4) in Zerilli 1970).
For example, we may choose a gauge in which
$g_{01}=0$. However, for such a gauge we will have $g_{02} \ne 0$.
What is significant is that we have a new frame dragging effect
due to non-vanishing $N_{\Sigma}{}^{\mu}$ because of the noncircularity
of the spacetime, as pointed out at
the end of \S~\ref{sec:GS}~C.

\begin{figure}
\plotone{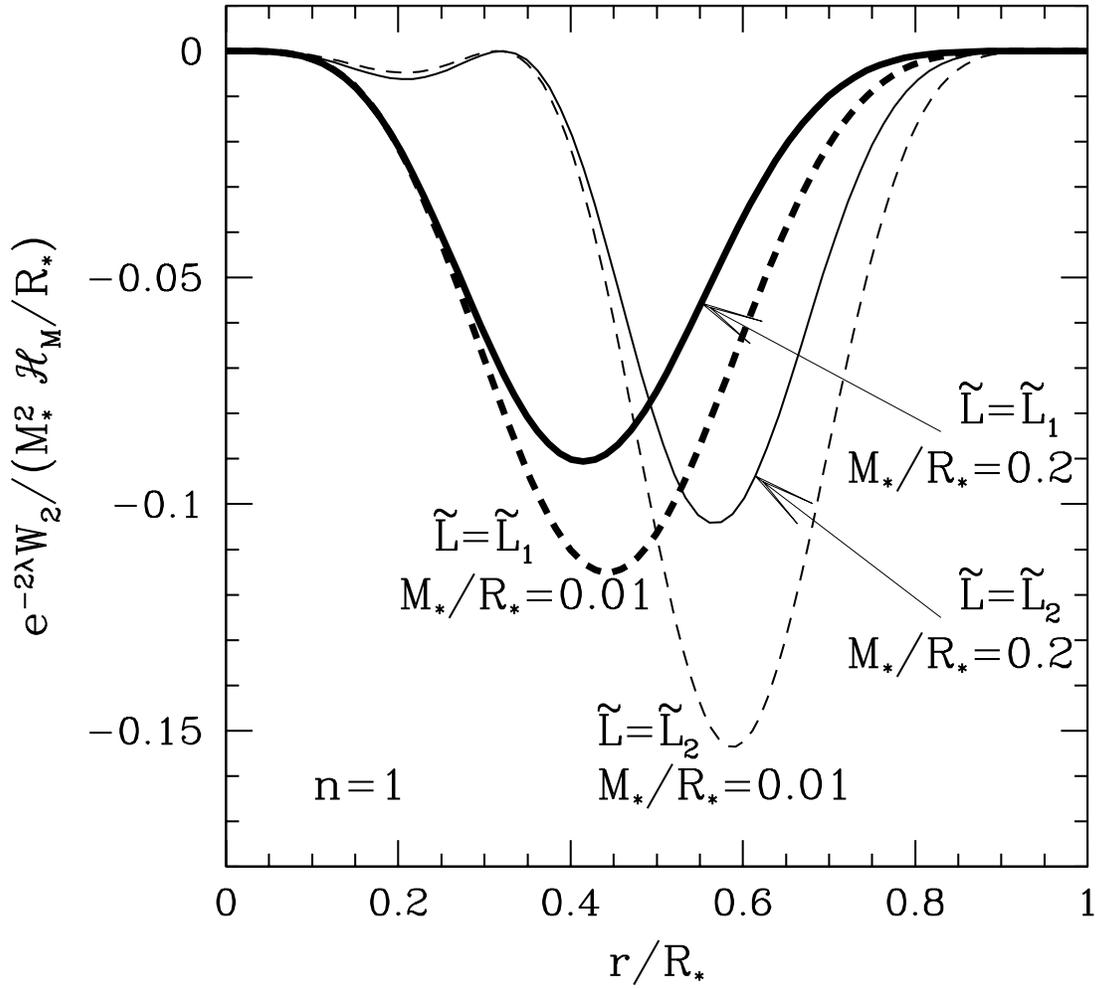}
\caption{\label{fig:w2} 
The $r\varphi$-component of the quadrupole ($\ell=2$) metric perturbation
as a function of radius, namely,
$e^{-2\lambda} W_{2}$ normalized by $M_{*}^2 {\cal H}_{M}/R_{*}$.
The solid and dashed lines are for $M_{*}/R_{*}=0.2$ (relativistic)
and $M_{*}/R_{*}=0.01$ (Newtonian), respectively, while
the bold and thin lines for $\tilde L=\tilde L_{1}$ 
and $\tilde L=\tilde L_{2}$ in Table~\ref{tab:eigen}, respectively.
We adopt the polytropic index $n=1$.
This metric component describes a new type of frame dragging due to
a toroidal magnetic field.
}
\end{figure}

In analogy with the above arguments,
let us consider the components of the
unit spacelike 4-vector $m^{\mu}$ orthogonal to the
$t=$const. and $\varphi=$const. hypersurface $\Sigma_{t\varphi}$
in equation (\ref{eq:defm}).
To the lowest order in $\Psi$,
from equations (\ref{eq:mcom}), (\ref{eq:metric}) and (\ref{eq:metricp}),
we have
\beqa
\frac{dr}{d\varphi}=\frac{m^{1}}{m^{3}}=M_{\Sigma}{}^{1}
=3 e^{-2\lambda} W_{2} \sin^2\theta \cos\theta\,.
\eeqa
In Figure \ref{fig:w2}, we show the quadrupole component of the
(normalized) coordinate derivative
$e^{-2\lambda} W_{2}/(M_{*}^2{\cal H}_{M}/R_{*})$
for $M_{*}/R_{*}=0.2$ and $M_{*}/R_{*}=0.01$.
We find from this figure that ${dr}/{d\varphi}\propto M_{*}/R_{*}$
for fixed mass and total magnetic helicity.
This `frame dragging' effect originates from the magnetic field
rather than the fluid flow,
since the right-hand side of equation (\ref{eq:W2})
does not depend on $\tilde C$ (or ${\cal R}_{V}$).
The component $W_{2}$ also violates the reflection symmetry about
the equatorial plane, i.e., $W_{2}$ is of parity $(-1)$ in
 equation (\ref{eq:metricp}).
Note, again, that the $r\varphi$-component ($W_{\ell}$) is gauge-specific
(see equation (D3) in Zerilli 1970).
For example, we can take a gauge in which $g_{13}=0$.
However, once again, an important point is that
we have non-vanishing $M_{\Sigma}{}^{\mu}$ because of the noncircularity,
which is a gauge-invariant statement. 

\section{Summary}\label{sec:sum}
We have investigated stationary axisymmetric configurations 
of magnetized stars in the framework of general relativistic ideal MHD.
A toroidal magnetic field and meridional flow in addition to 
a poloidal magnetic field are incorporated into 
the relativistic stellar model for the first time.
We have treated the magnetic field and meridional flow as perturbations,
but no other approximations were made.
We have restricted ourselves to a star with
a dipole magnetic field, slow rotation and polytropic equation of state,
and considered magnetic field configurations confined in the
interior of the star.

The solutions are found to be characterized by discrete eigenvalues
that describe the radial profiles of the magnetic field and fluid flow.
Namely, the magnetic field and velocity field configurations
become more complicated for larger eigenvalues.
We have found that the stellar shape is prolate rather than oblate
because of the stress of the toroidal magnetic field.
For fixed total baryonic mass and magnetic helicity,
more spherical stars are found to have lower energy. 
We have also found two new types of frame dragging
that differ from the one in rotating stars and Kerr black holes.
These effects are due to the presence of the meridional flow and 
toroidal magnetic field.
Interestingly, these effects violate the reflection symmetry 
about the equatorial plane.
This means that the magnetic field or the meridional flow
naturally selects the preferred direction about the axis of symmetry.
This mechanism may be related to the neutron star kick
or the jet formation.

\acknowledgments
KI is grateful to F.~Takahara and H.~Nakano for useful discussions.
This work was supported in part by the
Monbukagaku-sho Grant-in-Aid for Scientific Research,
Nos.~00660 (KI) and 14047214 (MS).



%\appendix

%
% Tables
%

\newpage
\clearpage
\begin{deluxetable}{ccccc}
\footnotesize
\tablecaption{\label{tab:eigen}
The first few roots of the (dimensionless) eigenvalue $R_{*} \tilde L$
in equation (\ref{eq:psi1}) are shown for several relativistic factors
$M_{*}/R_{*}$.
Note that the configuration $M_{*}/R_{*}=0.24$ is unstable
against radial perturbations.
}
\tablecolumns{5}
\tablewidth{0pc}
\tablehead{
\colhead{} & 
\colhead{$M_{*}/R_{*}=0.24$} & 
\colhead{$M_{*}/R_{*}=0.2$} &
\colhead{$M_{*}/R_{*}=0.1$} & 
\colhead{$M_{*}/R_{*}=0.01$}}
\startdata
$|R_{*} \tilde L_{1}|$ & 2.8704 & 3.9079 & 5.7924 & 7.2638 \\
$|R_{*} \tilde L_{2}|$ & 4.1284 & 5.6017 & 8.3151 & 10.475 \\
$|R_{*} \tilde L_{3}|$ & 5.3393 & 7.2568 & 10.795 & 13.615 \\
$|R_{*} \tilde L_{4}|$ & 6.5432 & 8.9031 & 13.258 & 16.729 \\
$|R_{*} \tilde L_{5}|$ & 7.7448 & 10.545 & 15.711 & 19.829 \\
\enddata
%\tablecomments{The uncertainties in the last
%quoted digits are given in parenthesis.}
%\tablenotetext{a}{calculated from ecliptic coordinates}
%\tablenotetext{b}{quoted value is median of asymmetric distribution}
\end{deluxetable}

%\newpage
%\clearpage
\begin{deluxetable}{ccrrrrrrrr}
%\rotate
\tabletypesize{\footnotesize}
\tablecaption{\label{tab:integ}
\footnotesize
Parameters of weakly magnetized stars in general relativity are shown
for several relativistic factors $M_{*}/R_{*}$ and eigenvalues
$\tilde L$ given in Table~\ref{tab:eigen}.
Shown are $(\Delta \rho)_{0c}$ (central density shift),
$\Delta M_{*}$ (mass shift),
$\Delta J$ (angular momentum),
$\Delta Q$ (mass quadrupole moment),
$\Delta V$ (current hexapole moment),
$(\Delta r)_{0}$ ($\ell=0$ deformation of the stellar surface),
$e_{*}$ (ellipticity at the stellar surface),
and ${\cal H}_{M}$ (dimensionless magnetic helicity).
The dimensionless quantities ${\cal R}_{V}$ and ${\cal R}_{M}$
defined in equations (\ref{eq:Rv}) and (\ref{eq:Rm}), respectively,
measure the magnitude of fluid flow and magnetic field energy.
The integer suffixed to each entry denotes
the powers of $10$ by which the entry number must be multiplied
(e.g., $1.1_{-2}$ means $1.1\times10^{-2}$).
}
\tablecolumns{10}
\tablewidth{0pc}
\tablehead{
\colhead{$M_{*}/R_{*}$} & 
\colhead{$\tilde L$} & 
\colhead{$\displaystyle \vphantom{\frac{\frac{f}{g}}{\frac{f}{g}}}
\frac{(\Delta \rho)_{0c}}{\rho_{c} {\cal H}_{M}}$} & 
\colhead{$\displaystyle\frac{\Delta M_{*}}{M_{*} {\cal H}_{M}}$} & 
\colhead{$\displaystyle\frac{\Delta J}{M_{*} R_{*} {\cal R}_{V}}$} & 
\colhead{$\displaystyle\frac{\Delta Q}{M_{*} R_{*}^2 {\cal H}_{M}}$} & 
\colhead{$\displaystyle\frac{\Delta V}{M_{*} R_{*}^3 {\cal R}_{V}}$} & 
\colhead{$\displaystyle\frac{(\Delta r)_{0}}{R_{*} {\cal H}_{M}}$} &
\colhead{$\displaystyle\frac{e_{*}}{{\cal H}_{M}}$} & 
\colhead{$\displaystyle\frac{{\cal H}_{M}}{{\cal R}_{M}}$} }
\startdata
0.24 & $\tilde L_{1}$ & 
$-1.7201_{-1}$ & $2.2123_{-2}$ & $-1.9030_{-2}$ & $-2.4813_{-3}$ & 
$1.6221_{-4}$ & $7.8089_{-2}$ & $-6.8901_{-3}$ & $1.8450_{-1}$ \\
$(\rho_{c} M_{*}^2=1.3304_{-2})$ & $\tilde L_{2}$ & 
$-1.3834_{-1}$ & $2.8090_{-2}$ & $-3.7484_{-2}$ & $-7.5644_{-3}$ & 
$5.0828_{-4}$ & $1.2146_{-1}$ & $-2.1005_{-2}$ & $2.2614_{-1}$ \\
$(p_{c}/\rho_{c}=5.4344_{-1})$ & $\tilde L_{3}$ & 
$-1.3433_{-1}$ & $3.3088_{-2}$ & $-1.3583_{-2}$ & $-1.3770_{-2}$ & 
$2.3390_{-4}$ & $1.6951_{-1}$ & $-3.8237_{-2}$ & $7.4362_{-2}$ \\
($\epsilon_{c}=n p_{c}/\rho_{c}$) & $\tilde L_{4}$ &
$-1.2457_{-1}$ & $3.8033_{-2}$ & $-1.7557_{-2}$ & $-2.0184_{-2}$ & 
$3.3273_{-4}$ & $2.1765_{-1}$ & $-5.6046_{-2}$ & $9.2887_{-2}$ \\
 & $\tilde L_{5}$ & 
$-1.2618_{-1}$ & $4.3057_{-2}$ & $-9.0243_{-3}$ & $-2.6556_{-2}$ & 
$1.7827_{-4}$ & $2.6526_{-1}$ & $-7.3741_{-2}$ & $4.7016_{-2}$ \\
\hline
0.2 & $\tilde L_{1}$ & 
$2.5443_{-1}$ & $2.4450_{-2}$ & $-2.5850_{-2}$ & $-6.1618_{-3}$ & 
$3.6056_{-4}$ & $-2.5991_{-2}$ & $-1.5050_{-2}$ & $2.5561_{-1}$ \\
$(\rho_{c} M_{*}^2=6.7418_{-3})$ & $\tilde L_{2}$ & 
$4.0620_{-1}$ & $3.0662_{-2}$ & $-7.0990_{-2}$ & $-1.8749_{-2}$ & 
$1.4779_{-3}$ & $9.0360_{-3}$ & $-4.5796_{-2}$ & $4.3463_{-1}$ \\
$(p_{c}/\rho_{c}=2.5582_{-1})$ & $\tilde L_{3}$ &
$5.3048_{-1}$ & $3.6320_{-2}$ & $-2.0397_{-2}$ & $-3.2930_{-2}$ &  
$5.1754_{-4}$ & $4.1634_{-2}$ & $-8.0431_{-2}$ & $1.1531_{-1}$ \\
($\epsilon_{c}=n p_{c}/\rho_{c}$) & $\tilde L_{4}$ & 
$6.7339_{-1}$ & $4.2004_{-2}$ & $-3.0716_{-2}$ & $-4.7192_{-2}$ &  
$8.3493_{-4}$ & $7.3899_{-2}$ & $-1.1527_{-1}$ & $1.6900_{-1}$ \\
 & $\tilde L_{5}$ & 
$7.9653_{-1}$ & $4.7794_{-2}$ & $-1.4039_{-2}$ & $-6.1198_{-2}$ &  
$3.9211_{-4}$ & $1.0519_{-1}$ & $-1.4948_{-1}$ & $7.6323_{-2}$ \\
\hline
0.1 & $\tilde L_{1}$ & 
$-4.2056_{-2}$ & $1.7802_{-2}$ & $-2.5913_{-2}$ & $-1.9571_{-2}$ &
$6.8215_{-4}$ & $9.0505_{-2}$ & $-3.6533_{-2}$ & $2.7778_{-1}$  \\
$(\rho_{c} M_{*}^2=7.9326_{-4})$ & $\tilde L_{2}$ & 
$1.1871_{-1}$ & $2.2276_{-2}$ & $-8.7908_{-2}$ & $-6.0171_{-2}$ &
$3.2927_{-3}$ & $1.8445_{-1}$ & $-1.1232_{-1}$ & $5.8027_{-1}$  \\
$(p_{c}/\rho_{c}=7.0273_{-2})$ & $\tilde L_{3}$ & 
$1.8560_{-1}$ & $2.6519_{-2}$ & $-2.2024_{-2}$ & $-1.0366_{-1}$ &
$9.7473_{-4}$ & $2.7967_{-1}$ & $-1.9350_{-1}$ & $1.3491_{-1}$  \\
$(\epsilon_{c}=n p_{c}/\rho_{c})$ & $\tilde L_{4}$ & 
$3.0621_{-1}$ & $3.0791_{-2}$ & $-3.6953_{-2}$ & $-1.4668_{-1}$ &
$1.7212_{-3}$ & $3.7327_{-1}$ & $-2.7381_{-1}$ & $2.2085_{-1}$  \\
 & $\tilde L_{5}$ & 
$3.8261_{-1}$ & $3.5149_{-2}$ & $-1.5645_{-2}$ & $-1.8861_{-1}$ &
$7.4237_{-4}$ & $4.6484_{-1}$ & $-3.5208_{-1}$ & $9.2516_{-2}$  \\
\hline
0.01 & $\tilde L_{1}$ & 
$-1.0732_{-1}$ & $2.2215_{-3}$ & $-2.3000_{-2}$ & $-3.9173_{-2}$ & 
$8.9803_{-4}$ & $1.3237_{-1}$ & $-5.9957_{-2}$ & $2.6217_{-1}$\\
$(\rho_{c} M_{*}^2=7.8563_{-7})$ & $\tilde L_{2}$ & 
$1.1140_{-1}$ & $2.7927_{-3}$ & $-7.9968_{-2}$ & $-1.2203_{-1}$ & 
$4.3637_{-3}$ & $2.7151_{-1}$ & $-1.8677_{-1}$ & $5.5877_{-1}$ \\
$(p_{c}/\rho_{c}=5.1447_{-3})$ & $\tilde L_{3}$ &
$1.8112_{-1}$ & $3.3291_{-3}$ & $-2.0174_{-2}$ & $-2.0944_{-1}$ & 
$1.2846_{-3}$ & $4.1207_{-1}$ & $-3.2056_{-1}$ & $1.3056_{-1}$ \\
$(\epsilon_{c}=n p_{c}/\rho_{c})$ & $\tilde L_{4}$ & 
$3.4219_{-1}$ & $3.8697_{-3}$ & $-3.4439_{-2}$ & $-2.9532_{-1}$ & 
$2.2933_{-3}$ & $5.4972_{-1}$ & $-4.5201_{-1}$ & $2.1748_{-1}$ \\
 & $\tilde L_{5}$ & 
$4.3170_{-1}$ & $4.4214_{-3}$ & $-1.4475_{-2}$ & $-3.7878_{-1}$ & 
$9.7915_{-4}$ & $6.8415_{-1}$ & $-5.7975_{-1}$ & $9.0471_{-2}$ \\
\enddata
%\tablecomments{The uncertainties in the last
%quoted digits are given in parenthesis.}
%\tablenotetext{a}{calculated from ecliptic coordinates}
%\tablenotetext{b}{quoted value is median of asymmetric distribution}
\end{deluxetable}

%
% Figure captions
%

\newpage
\clearpage

\end{document}